\documentclass[useAMS,usenatbib,fleqn]{mn2e}
\usepackage[pdftex]{graphicx}
\usepackage{amsmath,amssymb,hyperref}
\usepackage{times}
\usepackage[T1]{fontenc}

\title[Polarization leakage in epoch of reionization windows]{Polarization leakage in epoch of reionization windows -- II. Primary beam model and direction dependent calibration}
\author[K. M. B. Asad et al.]
{K. M. B. Asad$^1$\thanks{E-mail: khan@astro.rug.nl},
L. V. E. Koopmans$^{1}$,
V. Jeli\'{c}$^{1,2,3}$,
A. Ghosh$^1$,
F. B. Abdalla$^4$,
\newauthor
M. A. Brentjens$^3$,
A. G. de Bruyn$^{1,3}$,
B. Ciardi$^5$,
B. K. Gehlot$^1$,
I. T. Iliev$^6$,
\newauthor
M. Mevius$^{1,3}$,
V. N. Pandey$^3$,
S. Yatawatta$^{1,3}$
and S. Zaroubi$^1$\\ \\
$^1$Kapteyn Astronomical Institute, University of Groningen, PO Box 800, NL-9700 AV Groningen, the Netherlands\\
$^2$Ru{\dj}er Bo\v{s}kovi\'{c} Institute, Bijeni\v{c}ka cesta 54, 10000 Zagreb, Croatia\\
$^3$ASTRON, PO Box 2, NL-7990 AA Dwingeloo, the Netherlands\\
$^4$Department of Physics \& Astronomy, University College London, Gower Street, London WC1E 6BT, UK\\
$^5$Max-Planck Institute for Astrophysics, Karl-Schwarzschild-Strasse 1, D-85748 Garching bei M\"{u}nchen, Germany\\
$^6$Astronomy Centre, Department of Physics \& Astronomy, Peven
sey II Building, University of Sussex, Falmer, Brighton BN1
9QH, UK\\
}

\begin{document}

\label{firstpage}
\date{Accepted . Received ; in original form ; submitted to MNRAS on 15 April 2016 }
\pagerange{\pageref{firstpage}--\pageref{lastpage}} \pubyear{2016}

\maketitle

\begin{abstract}
Leakage of diffuse polarized emission into Stokes $I$ caused by the polarized primary beam of the instrument might mimic the spectral structure of the 21-cm signal coming from the epoch of reionization (EoR) making their separation difficult.
Therefore, understanding polarimetric performance of the antenna is crucial for a successful detection of the EoR signal.
Here, we have calculated the accuracy of the nominal model beam of LOFAR in predicting the leakage from Stokes $I$ to $Q,U$ by comparing them with the corresponding leakage of compact sources actually observed in the 3C295 field.
We have found that the model beam has errors of $\le 10\%$ on the predicted levels of leakage of $\sim 1\%$ within the field of view, i. e. if the leakage is taken out perfectly using this model the leakage will reduce to $10^{-3}$ of the Stokes $I$ flux.
If similar levels of accuracy can be obtained in removing leakage from Stokes $Q,U$ to $I$, we can say, based on the results of our previous paper, that the removal of this leakage using this beam model would ensure that the leakage is well below the expected EoR signal in almost the whole instrumental $k$-space of the cylindrical power spectrum.
We have also shown here that direction dependent calibration can remove instrumentally polarized compact sources, given an unpolarized sky model, very close to the local noise level.
\end{abstract}

\begin{keywords}
polarization -- techniques: polarimetric -- dark ages, reionization, first stars
\end{keywords}

\section{Introduction}
One of the fundamental obstacles in statistically detecting the 21-cm signal coming from the epoch of reionization (EoR) is the leakage of polarized signal into total intensity caused by the time-frequency-baseline-direction dependent primary beams of the telescope.
The Galactic diffuse foreground, the most dominant contaminant of the EoR signal after the extragalactic compact sources \citep[e.g.][]{be09,be10,pa14}, is expected to be separated from the signal by utilizing the fact that the foreground is spectrally smooth and the signal is not \citep{je08,da10,ha10,tr12,mo12,be13,po13,ch13,di15,th15}.
However, the Faraday-rotated polarized foreground is also not smooth along frequency, and its leakage into total intensity might mimic the frequency structure of the EoR signal making the separation of the two difficult \citep[hereafter A15]{pe09,je10,mo13,a}.
Moreover, chromaticity of the beam---characterized, e. g., by the first derivative of the beam as a function of frequency---can cause the spectrally smooth diffuse foreground to show fluctuations along frequency \citep{mo}.
The EoR detection experiments with 
GMRT\footnote{http://gmrt.ncra.tifr.res.in/} \citep{pe09},
LOFAR\footnote{http://www.lofar.org/} (A15, \citealt{je15}),
PAPER\footnote{http://eor.berkeley.edu/} \citep{ko16},
MWA\footnote{http://www.mwatelescope.org/} \citep{su15},
HERA\footnote{http://reionization.org/} \citep{ne16a},
SKA\footnote{http://www.skatelescope.org/} \citep{le15} will be affected by `polarization leakage' to various degrees depending on the directional gain properties and the fields of view of the instruments.

A number of papers dealing with the direction dependent (DD) gains, i. e. the primary beam, of low frequency radio telescopes have been published recently that shows the relevance of polarimetric analysis in the detection of the EoR signal.
\citet{po16} demonstrated the effects of the Stokes $I$ primary beam of MWA that can leak power from the foreground wedge into the EoR window, claiming that the foreground in even the sidelobes of the primary beam needs to be modeled and removed for a successful detection of the EoR signal as the farther the source is from the phase center the worse the leakage of power from the wedge.
Polarized power will be leaked from the wedge in a similar way albeit to a much lower amplitude.
Efforts are underway to better understand the MWA beam and calculate the accuracy of the beam model.
\citet{su15} found that with the `Full Embedded Element Pattern' model, the beam can be 2--5\% different from reality.
With the improved model they found that a $I\rightarrow Q$ leakage of a few per cent (with outliers up to 10\%) is achievable which is higher than that of the LOFAR case, as the field of view of an MWA tile is significantly higher than that of a LOFAR station.
Some preliminary results of the `intrinsic cross-polarization ratio' \citep[IXR;][]{cw} of MWA tiles \citep{su13} have been published.
\citet[fig. 1]{fo15} show an example of the variation of IXR$_J$ ($IXR$ calculated in terms of Jones matrices; see \citet{cw} for more details) of a simple all sky dipole element at 130 MHz.
The beam model exhibits an IXR$_J$ of 70 dB toward the zenith with a low-IXR$_J$ structure along the 45 degree line between the orthogonal receptors, and based on this model they have used a polarization leakage of up to -30 dB in their simulations of pulsar times of arrival.
A stringent limit on the accuracy of the model beamwidth of a  wide-field transit radio telescope has been set by \citet{sh15}.
By simulating the CHIME\footnote{http://chime.phas.ubc.ca} observations of the foreground-contaminated 21-cm signal in the presence of instrumental errors, they have found that in order to recover unbiased power spectra, the model beamwidth of each element should be known to an accuracy of at least 0.1\% within each minute.
Compared to beamwidth, beamshape errors would be even more difficult to model and hence would be a more problematic source of bias in the power spectra.

In a previous paper (A15), we predicted the polarization leakage from Stokes $Q,U$ to $I$ to be expected in the `EoR window' of the cylindrically averaged power spectra (PS) using the LOFAR observations of the 3C196 field.
The prediction was based on the nominal model beam of LOFAR produced by \citet{ha11} using an electromagnetic simulation of the ASTRON Antenna Group\footnote{M. J. Arts; http://www.astron.nl} \citep{sn07}.
We found that within a field of view (FoV) of 3 degrees the rms of the leakage as a fraction of the rms of the polarized emission varies between 0.2 to 0.3 per cent, and the leakage is lower than the EoR signal at $k<0.3$ Mpc$^{-1}$.
We thus concluded that even a modest polarimetric calibration over the FoV would ensure that the polarization leakage remains well below the expected EoR signal at scales of 0.02--1 Mpc$^{-1}$.
The accuracy of this prediction depends mainly on the accuracy of the model beam.
\begin{figure}
\centering
\includegraphics[width=\linewidth]{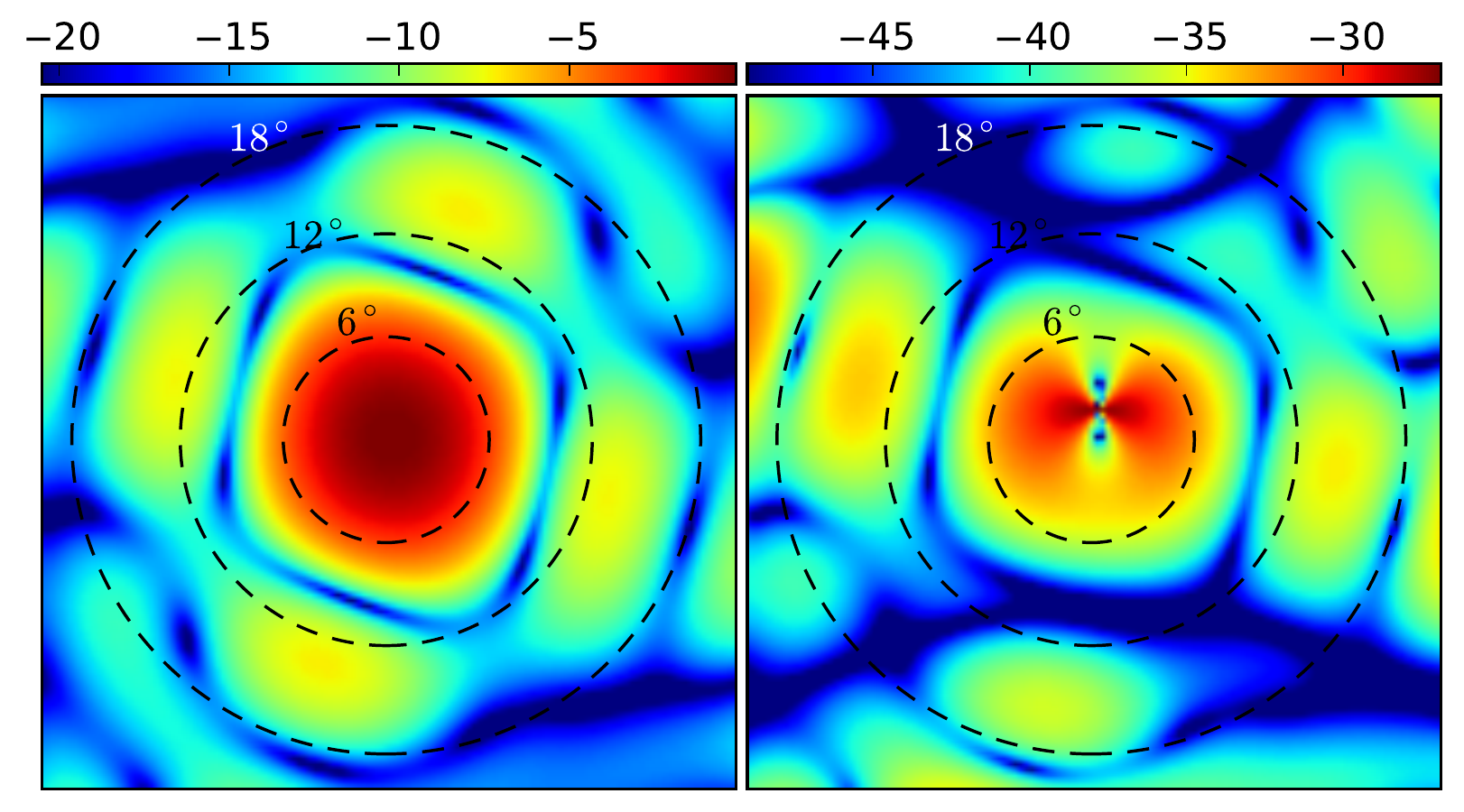}
\caption{Simulated model of the primary beam Jones matrix of the LOFAR station CS001HBA0 at 150 MHz for a zenith pointing. The color-bar is shown in decibel units. The left and right panels show the $(xx+yy)/2$ and $(xx-yy)/2$ components respectively.}
\label{f:beam}
\end{figure}
\begin{figure}
\centering
\includegraphics[width=\linewidth]{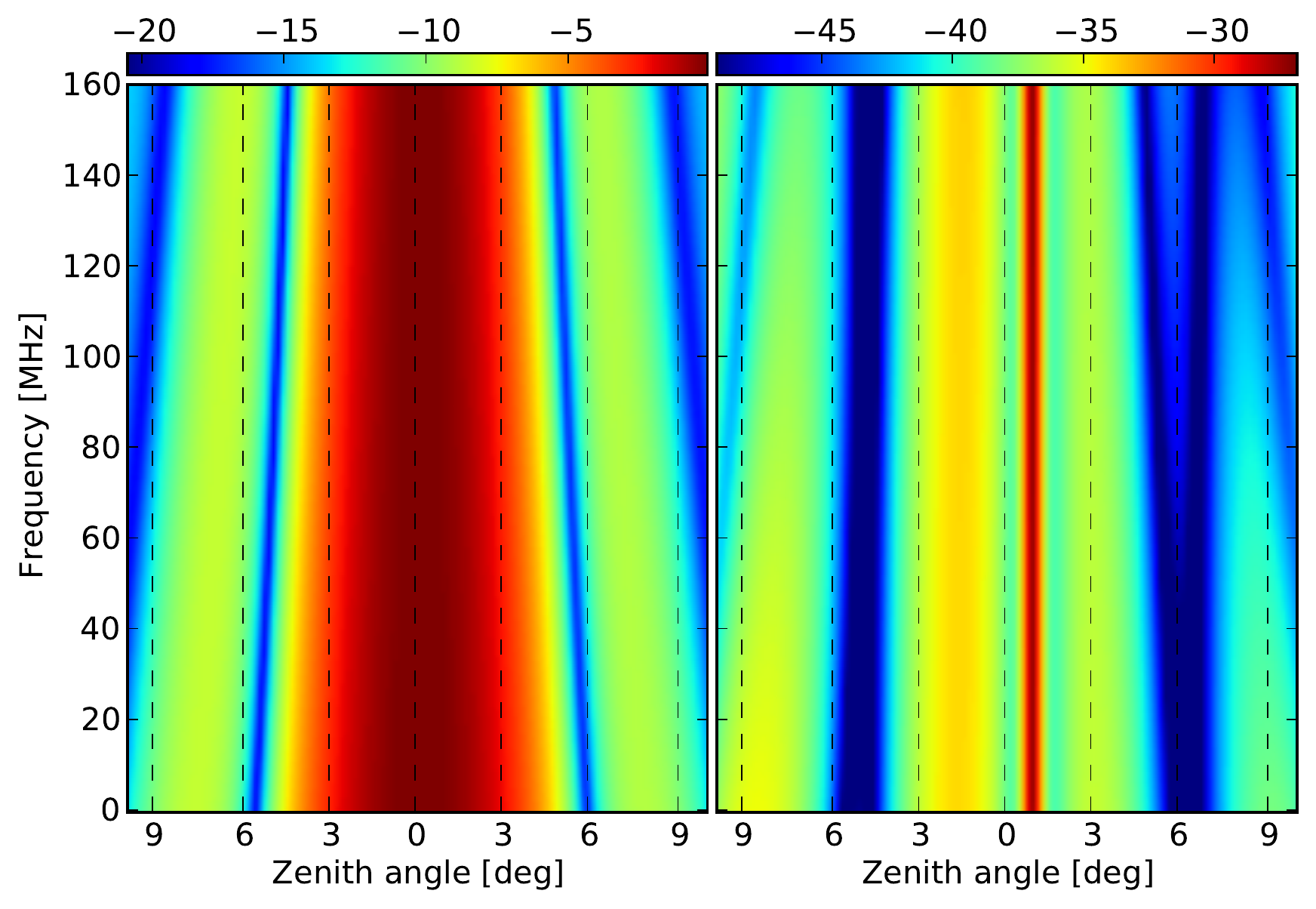}
\caption{Spectral structure of the primary beam model of LOFAR calculated by taking a slice in frequency space at different zenith angles at an azimuth of 180$^\circ$. The color-bar is shown in decibel units. The left and right panels show the $(xx+yy)/2$ and $(xx-yy)/2$ components respectively.}
\label{f:beamnu}
\end{figure}
In the current paper, we have used LOFAR observations of the compact sources in the 3C295 field to quantify the accuracy of the nominal model beam of LOFAR \citep{ha11}, as this field is less contaminated by polarized diffuse emission than the 3C196 field.
In addition to quantifying the accuracy of the beam, we demonstrate the efficiency of a DD calibration method in removing instrumentally polarized compact sources.
This paper is organized as follows.
Section 2 revisits the nominal model beam of LOFAR and shows the behavior of the intrinsic cross-polarization ratio of the instrument as a function of distance from the phase center and also distance of the observing field from the zenith.
In section 3, we describe the data reduction, calibration and simulation pipelines.
Our results are presented in section 4---first, we present the results of the observation and the simulation, then compare them to quantify the accuracy of the beam model, and finally present the results of the DD calibration.
The paper ends with a discussion of our analysis and some concluding remarks.
\begin{figure*}
\centering
\includegraphics[width=0.9\textwidth]{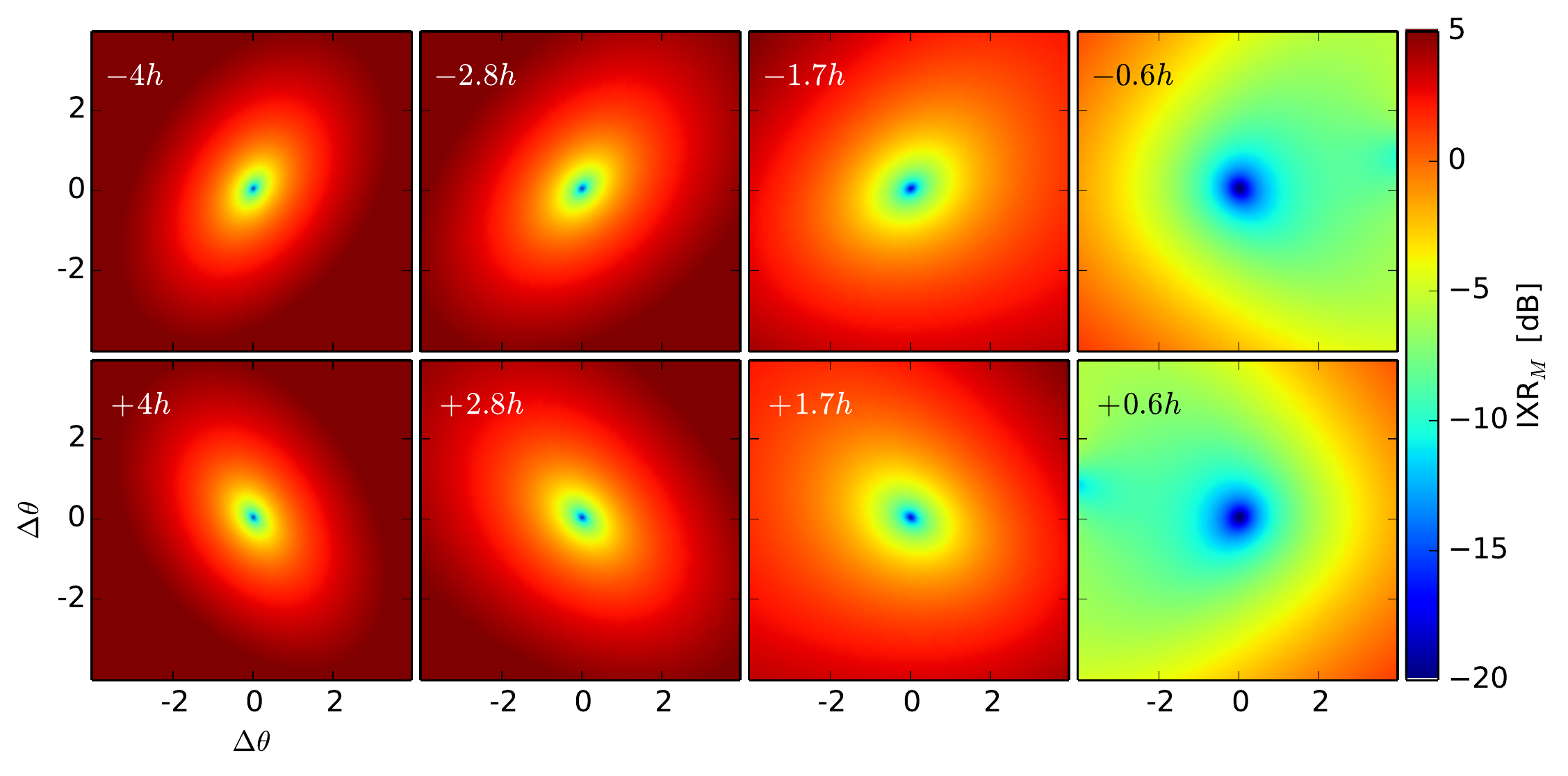}
\caption{IXR$_M$ of a typical LOFAR baseline within the central $8.3^\circ\times 8.3^\circ$ of the 3C295 field at eight different instances during an 8-hr synthesis. The EM-simulated LOFAR model beam has been used to calculate the parameter. The panels correspond to different hour angles, mentioned in the top-left corners. IXR$_M$ is lowest at $\mp$0.6 h when the field is very close to its culmination point or equivalently in the zenith.}
\label{f:ixr2}
\end{figure*}
\section{Primary beam model of LOFAR}
The primary beams of LOFAR HBA stations are modeled in three steps: an analytic expression is used for the dipole beams whose coefficients are calculated by fitting to a beam raster generated by electromagnetic simulation, the 16 dipole beams are phased in an analog way to create the tile beams, and the tile beam patterns are multiplied together with the respective weights and phases using an `array factor' to create the station beams.\footnote{The simulation procedure is described in detail in \citet{ha11} and its key points are also mentioned in \citet[section 2.2.2]{a}. Also note that the software package that creates the directional response of the antenna elements is included in the standard LOFAR calibration software BBS \citep{pa09}, publicly available at \url{https://svn.astron.nl/LOFAR/trunk/CEP/Calibration}.}
As a low-frequency aperture array, LOFAR does not have any moving parts and hence the beams of two orthogonal feeds are projected non-orthogonally away from zenith while tracking a moving source giving rise to mutual coupling between the beams.
The projection-induced mutual coupling is the principal contributor to polarization leakage and its removal completely depends on how well we can model the beam.

\begin{figure}
\includegraphics[width=\linewidth]{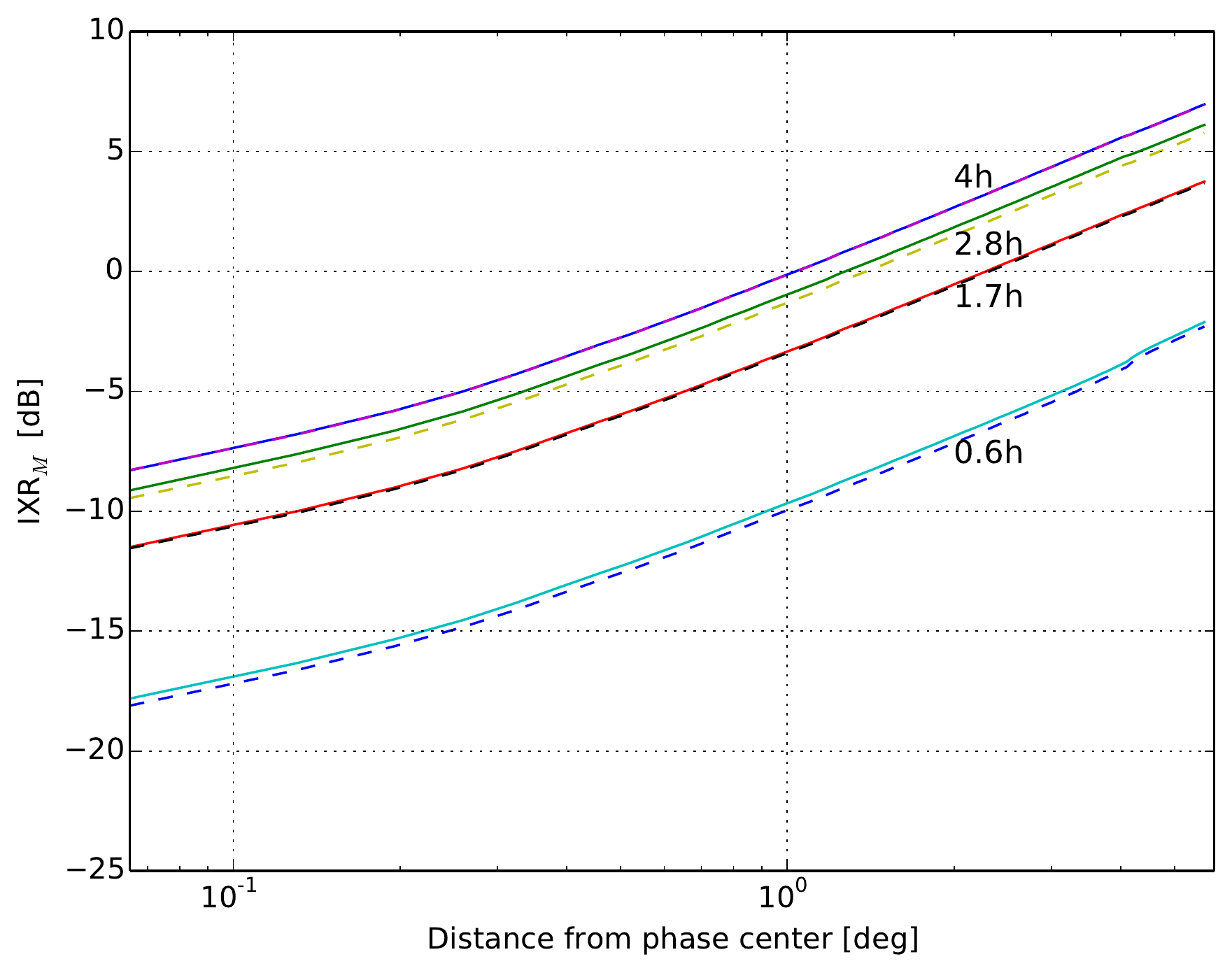}
\caption{Azimuthal profiles of IXR$_M$ at eight different instances of time during an eight hour observation of the 3C295 field. The texts correspond to the hour angle of the field during the observation which clearly shows that the leakage is lowest at $\pm$0.6 h. The solid and dashed lines correspond to the negative and positive hour angles respectively.}
\label{f:ixr1}
\end{figure}

Although projection effects are worse at lower elevation, it is possible to remove the effects of the primary beam to high accuracy toward at least one direction, the phase center, at all elevations using direction independent (DI) calibration \citep{sa96} and a model of the dipole beam (e. g. A15). After calculating the electronic gains of a station via DI calibration and correcting the data for the dipole beam at the phase center, only the effect of the array factor and the errors in correcting the dipole beam effects remain, which can be thought of as a differential beam with respect to the phase center. An example of such a `differential'\footnote{In this paper, whenever we talk about the LOFAR primary beam or station beam, it should be understood to be the `differential' primary beam, i. e. the beam where each component of the Jones or Mueller matrix has been normalized with respect to the phase center.} station beam at 150 MHz is shown in Fig. \ref{f:beam} where the field of view, the nulls and the sidelobes are clearly visible.
The left panel of the figure shows the sum of the diagonal terms of the beam Jones matrix, i. e. $(xx+yy)/2$, and the right panel shows their difference, i. e. $(xx-yy)/2$.
If we divide the difference by the sum, we obtain the fraction of Stokes $I$ flux that leaks into Stokes $Q$.
A more intuitive way to calculate this leakage is to use Mueller matrices instead of Jones matrices, and we calculate the leakage in terms of Mueller matrices below and follow the Mueller formalism throughout the paper.
Fig. \ref{f:beamnu} shows the spectral structure of the sum (left panel) and the difference (right panel) of the diagonal terms of the beam Jones matrix and they demonstrate that the position of the sidelobes changes \textit{smoothly} as a function of frequency.
We will demonstrate the accuracy of this beam in predicting the polarization leakage. We will call this leakage the `off-axis' leakage as opposed to the `on-axis' leakage at the phase center. Off-axis leakage increases as we go away from the phase center and the zenith or, in case the observing field never reaches the local zenith, the culmination point of the field.

A fundamental figure of merit (FoM) to evaluate the polarization performance of a polarimeter is the intrinsic cross-polarization ratio (IXR) introduced by \citet{cw}. The `intrinsic' in IXR signifies that the parameter is independent of the choice of coordinate systems. IXR is related to the invertibility of a DD Jones matrix. The Jones matrices calculated by calibration are inverted and multiplied with the data to give the `corrected' data, and hence the intrinsic invertibility of a Jones matrix put a fundamental limit to the extent to which a data can be corrected. For Stokes polarimeters, IXR can be easily converted to a Mueller IXR, or IXR$_M$ which, in turn, is directly related to the fractional polarization leakage (fraction of Stokes $I$ signal leaked into Stokes $Q,U,V$ and vice versa) caused by the beam, mathematically
\begin{equation}
\text{IXR}_M = 10\times \log_{10}\left[\frac{\sqrt{M_{10}^2+M_{20}^2+M_{30}^2}}{|M_{00}|}\right] \text{dB}
\label{eq:ixr}
\end{equation}
where $M$ is a $4\times 4$ Mueller matrix corresponding to the outer product of the DD Jones matrices of two elements that make up a baseline of an array, and by the subscript `10' in $M_{10}$ we mean the second row and first column of the matrix (for explanation see section 2.2.1 of A15). $M_{10}$, $M_{20}$, and $M_{30}$ give leakages from $I$ to $Q$, $U$, and $V$ respectively, and $M_{00}$ gives the Stokes $I$ beam. For an example of a Mueller matrix that completely characterizes the beam of a baseline see fig. 2 of A15. Note that IXR$_M$ is usually taken to be the opposite of this value, the values are usually expressed as a positive integer and in dB units. However, here we express the dB values as negative integers so that they correspond to the increment of leakage with distance from the phase center more intuitively.

IXR$_M$ distributions within the central $8.3^\circ\times 8.3^\circ$ of the 3C295 field, one of the secondary observing windows of the LOFAR-EoR project, at eight different instances of time during an 8 hour synthesis are shown in Fig. \ref{f:ixr2} for example; the observation time increases as we go from left to right panels on the top, and then from right to left panels on the bottom.
We see that IXR$_{M}$ increases as we go away from the phase center of the field and also from the culmination point, and this increment directly corresponds to an increase in leakage.
There is a reversal of orientation of the elliptical shape of the spatial distribution of the IXR$_M$ which is due to the reversal of the orientation of the projected dipole beams.
For a more quantitative understanding we show the azimuthal profiles of IXR$_M$ at the same eight instances of time in Fig. \ref{f:ixr1}. We see the same trend as Fig. \ref{f:ixr2} here: an increase of IXR$_M$ as we go away from the center and the zenith.
IXR$_M$ or equivalently the leakage is lowest near the zenith, 3.4 $\sim$ 4.5 hours after the beginning of the synthesis.

All plots in this section have been calculated from the model beam of LOFAR \citep{ha11}, the same beam that was used to predict the amount of leakage from linearly polarized diffuse Galactic emission into total intensity (A15). Our main aim in this paper is to demonstrate the accuracy of this model within the FoV of a typical LOFAR HBA (high band antenna, 110--200 MHz) station. After finding the accuracy, we will be able to constrain our previous prediction more robustly and the need for improvement of the model. The accuracy has been demonstrated below by comparing the leakage actually seen in an 8-hr synthesis data of the 3C295 field and the leakage predicted by the model beam that we have introduced in this section.

\section{Data processing and simulation pripelines}

We have used real and simulated LOFAR observations of the 3C295 field.
This field was chosen because the compact sources in Stokes $Q,U$ leaked from Stokes $I$ are less contaminated by diffuse polarized emission compared to the 3C196 field.
Observed data were processed using the standard LOFAR software pipeline, and the simulated observations were produced using the simulation pipeline presented in A15.
In this section, we briefly describe the observational setup and data processing steps. Next, the process of simulating the desired observations, taking into account the systematic effects of LOFAR, by implementing our previous pipeline is described.

\begin{table}
\centering
\begin{minipage}{\linewidth}
\centering
\caption{Observational parameters of the 3C295 synthesis.}
\label{t:obs}
\begin{tabular}{@{}lllr@{}}
\hline
\hline
 & 3C295 field \\
\hline
Observation ID & L104068 \\
Start time [UTC] & 22 Mar, 2013, 21:41:05 \\
Phase center, RA & 14$^h$11$^m$20.6$^s$ \\
Phase center, DEC & +52$^\circ$12'9'' \\
Frequency range & 115--189 MHz \\
Spectral resolution & 3.2 kHz \\
Observing time & 8h \\
Integration time & 2s \\
\hline
\end{tabular}
\end{minipage}
\end{table}

\subsection{Observations}
\begin{figure}
\includegraphics[width=\linewidth]{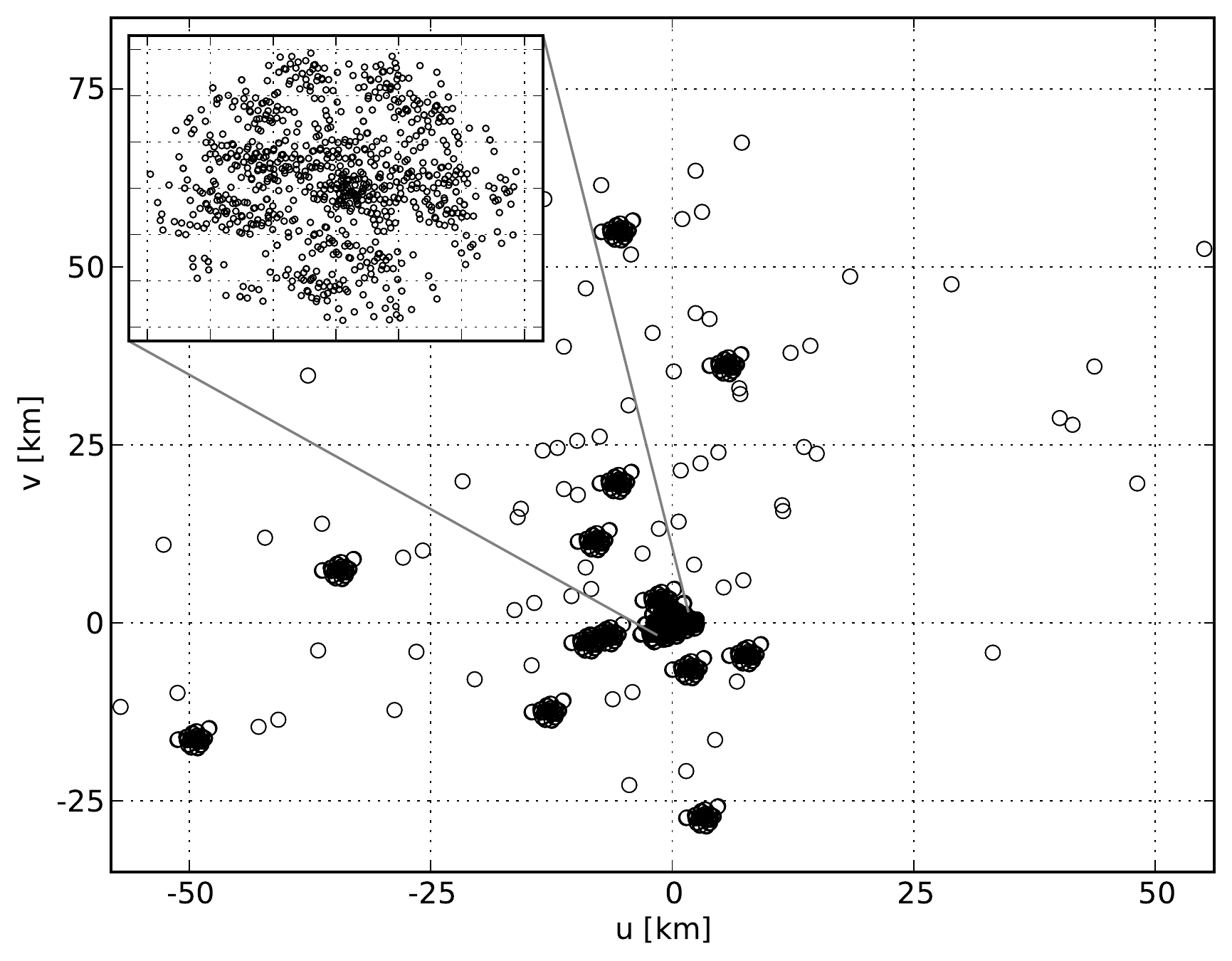}
\caption{The instantaneous uv-coverage of the LOFAR configuration. Only the Dutch stations have been shown here. The coverage within the inner 3km, relevant for our experiment, is shown on the inset. The $u$ and $v$ distances are shown in frequency independent physical units, i. e. in km.}
\label{f:uv}
\end{figure}
The 3C295 field was observed multiple times by LOFAR.
Here we have used an 8-hour synthesis observation taken in March 2013.
An overview of the observational parameters is presented in Table \ref{t:obs}, and an instantaneous uv-coverage of the inner 3 km of the configuration used for this observation is shown in Fig. \ref{f:uv}.

For this observation, the phased array was set up in the HBA DUAL INNER configuration consisting of 48 core stations (CS) and 14 remote stations (RS).
As the RS have 48 tiles in contrast to the 24-tile CS \citep[fig. 4]{vh}, half of the tiles of the RS were turned off to make them equivalent to the CS. The observations spanned the frequency range from 115 to 189 MHz, that was divided into 380 sub-bands, each of width $\sim 195$ kHz.
Each subband was further divided into 64 channels. All four correlations of the voltages between the orthogonal pairs of dipoles were recorded and the data were integrated every 2s at the correlator.
Data were taken only during the night and the syntheses were symmetric around the time of culmination of the fields.

\subsection{Flagging and averaging}

The acquired data were first processed using the {\tt AOFlagger} \citep{of10,of12} to remove terrestrial radio frequency interference (RFI).
Within the frequency range from 115--177 MHz, on average only 1\% data were flagged due to RFI.
However, above 177 MHz more than 40\% of the data were flagged due to interfering signals from Digital Audio Broadcasting.
After flagging, the data were averaged in time and frequency to reduce their volume for further processing.
Every 10s of the data were averaged, and the inner 60 channels of every sub-band were averaged to produce a single channel of width 183 kHz.
The four edge channels were excluded from the averaging process to remove edge effects from the polyphase filter.
Averaging usually results in bandwidth and time smearing, but we are not affected by them as only the short baselines were considered in this study which are less prone to smearing effects.

\subsection{Calibration}

After flagging and averaging, we performed calibration in two steps: direction independent (DI) and direction dependent (DD).
DI calibration was performed using the {\tt Black Board Selfcal} ({\tt BBS}) package \citep{pa09}.
We used a sky model consisting of only 3C295, the central source that dominates the visibilities on all baselines.
The model \citep{sc} had two components with a total Stokes $I$ flux of $97.76\pm 2$ Jy at 150 MHz and it also sets our broad band spectral model.
Note that the model "lead to unacceptably high flux scale uncertainty at frequencies below 70 MHz," but we would not be affected by this as we restrict ourselves within the frequency range of 134--166 MHz.
{\tt BBS} calculates the four complex components of the DI Jones matrices for each station taking into account the changing location of 3C295 within the primary beam of the dipole elements, and the variation of parallactic angle which minimizes the instrumental polarization in the vicinity of the phase center of the field.
The gains are then applied on the model visibilities which in turn are subtracted from the observed visibilities to remove the 3C295 source with its DI gains.
In addition, {\tt BBS} removes the clock and short-timescale ionospheric phase errors, and sets the frequency-dependent intensity and astrometric reference frame for the field \citep{pa09,ya12,je14}.

DI-calibration works well only for the sources on or very close to the phase center.
However, there are another 10 sources brighter than 0.75 Jy within the central $8^\circ$ of the 3C295 field, and we removed them with their corresponding DD-gains using {\tt SAGECAL}, a DD-calibration package \citep{ka11,ka13}.
{\tt SAGECAL} calculates complex Jones matrices for every station toward the directions of the 10 sources.
Although {\tt SAGECAL} does not have any information about the time-frequency varying polarized primary beams of the stations, it should be able to reproduce their effects through the DD-gains.
In principle, all significant DD effects should be absorbed in the DD-gains, among them also the position-dependent ionospheric delays.
Each direction is associated with one source, and the solution interval is 10 minutes for each direction which is sufficient to remove the sources down to the confusion noise, resulting from the background unresolved radio sources, on the short baselines.
Removing these 10 sources does not affect the other sources in the field since no gain solutions are applied to the data in DD calibration; they are only applied to the model and subtracted from the data.
Note that the data has not been corrected for ionospheric Faraday rotation, that depolarizes the signal depending on the level of total electron content in the ionosphere.
However, it will not affect our experiment as the variation of the ionospheric Faraday rotation is usually comparatively small within 8 hours \citep[e. g. see fig. 2 of][]{je15} and it affects only the intrinsically polarized sources which are excluded while calculating the accuracy of the beam model.

\begin{table}
\begin{minipage}{\linewidth}
\centering
\caption{Imaging parameters.}
\label{t:img}
\begin{tabular}{@{}lllr@{}}
\hline
\hline
Baseline cut & $30 - 1000 \ \lambda$ \\
Weighting & Natural \\
Angular resolution (PSF) & 3.44 arcmin \\
Frequency range & 134--166 MHz \\
Spectral resolution & 1.9 MHz \\
Synthesis time & 8h \\
Time resolution & 10s \\
Number of pixels in the image & 1024 $\times$ 1024 \\
Size of each pixel & 0.5 arcmin \\
\hline
\end{tabular}
\end{minipage}
\end{table}

\subsection{Imaging}

Two different sets of images were produced, one from only the DI-calibrated data and the other from both DI- and DD-calibrated data. Imaging was performed using the standard LOFAR-EoR imaging software, excon \citep[\url{http://exconimager.sf.net}]{ya14}.
Baselines only up to 1 k$\lambda$ were used and, although higher resolution images would produce even better results, an angular resolution of 3.44 arcmin is both sufficient for our purposes and computationally less expensive.
The visibilities were weighted naturally in all cases.
We took 160 subbands within the frequency range of 134--166 MHz centered around 150 MHz to conduct our simulation and analysis described below.
We use the same parameters to create images from both the real observations and the simulated observations. The imaging parameters are listed in Table \ref{t:img}.

\subsection{Simulated observations}

\begin{table}
\begin{minipage}{\linewidth}
\centering
\caption{Parameters for the simulated observation.}
\label{t:sim}
\begin{tabular}{@{}lllr@{}}
\hline
\hline
Baselines used & $30 - 1000 \ \lambda$ \\
Phase center, RA & 14$^h$11$^m$20.6$^s$ \\
Phase center, DEC & +52$^\circ$12'9'' \\
Frequency range & 134--166 MHz \\
Spectral resolution & 1.9 MHz \\
Synthesis time & 8h \\
Time resolution & 10s \\
Minimum flux in the sky model & 100 mJy \\
Number of sources & 140 \\
\hline
\end{tabular}
\end{minipage}
\end{table}

For simulating the LOFAR observations of the 3C295 field, we use the pipeline described in our previous paper, A15.
Here we briefly outline the steps of the simulation specific to this experiment.
\begin{enumerate}
\item First step in simulating an observation is to create a realistic sky model from observed data that we want to compare with. We have taken the Stokes $I$ images for 160 frequency channels and created a sky model using {\tt buildsky} that uses the available frequency information to calculate the spectral index of each source. The aim is to predict the leakage from Stokes $I$ to $Q,U$, hence we do not want to include any polarization in our sky model. We used a flux cut of 100 mJy to remain well above the local noise around the sources ($\sim 5$ mJy/beam). Our model consisted of 140 point sources, mainly within the first null of the primary beam, several of which were constructed by more than one components. Note that, we have created a sky model from an image that was not corrected for the primary beam. Therefore, there is a systematic decrease in flux away from the phase center until the first null, and then there are some more sources on the sidelobes of the beam. The attenuation caused by the primary beam does not pose any difficulty in quantifying the fractional leakages, as the attenuation effect drops out in the ratio of different Stokes parameters, the parameter we are interested in. Also note that in the case of calibrating real data, sky models are usually constructed from very high resolution images, but for our purpose such precision is not required as, again, we are only interested in the fraction of Stokes $I$ flux leaked into the other Stokes parameters, and not in the absolute Stokes $I$ flux.
\item We calculated visibilities from the sky model using the same baselines as that of the observation at all frequency channels and taking into account the station-time-frequency dependent model primary beams of the instrument. This was done, in effect, by multiplying the fluxes of the individual sources with the beams at the corresponding positions, times and frequencies and Fourier transforming them using {\tt BBS}. Therefore, although the sky was completely unpolarized, the predicted visibilities had non-zero values in all four visibility correlations due to instrumental polarization. The parameters of this simulated observation are listed in Table \ref{t:sim}. Note that, the set-up of the instrument was the same for the simulation as that of the observation.
\item The simulated visibilities were inverted to produce four images corresponding to the four Stokes parameters. Same imaging parameters were used in this case as in the case of imaging from the observed data; the key parameters are listed in Table \ref{t:img}. The standard imaging software {\tt CASA} was used for all imaging. Although we created images for all Stokes parameters, here we will use only Stokes $I$, $Q$ and $U$ images for our analysis, as the SNR of $I\rightarrow V$ leakage is too low to be useful for a comparison between the observation and the simulation. We used the standard definition of Stokes visibilities to calculate the Stokes parameters from the visibility correlations, as given by the equations (13)a--d of A15, and the linear polarization $P$ was calculated as $Q+iU$.
We also created an average of the images of all frequency channels to get an increased SNR that will facilitate the extraction of source fluxes.
\end{enumerate}

\subsection{Source flux extraction}

Once we have all the images, the next step was to calculate the fluxes of the sources that we want to compare.
The quantity we use in this case is the `peak flux', as it is straightforward to determine and sufficient for illustrating the difference between the observation and the simulation.
To extract the peak fluxes, first, we created small non-overlapping circular apertures around the point sources of interest in the averaged observed $P$ image---as the source must be present in the observed $P$ image for us to be able to compare it with the simulation---where the sources are clearly visible due to high SNR.
The sizes of the apertures depended on the structure of the sources, some of which had double lobes, but their radii were never more than 3.3 arcmin in an 8.3 degree image of 0.5 arcmin pixel size.
The apertures thus produced from the observed $P$ image were used in all images, and the maximum flux within the apertures were extracted in each case.
Following this method, we produced eight different lists of the sources with their peak flux corresponding to the Stokes $I$, $Q$, $U$ and the linear polarization $P$ images of both the observation and the simulation.
The minimum threshold set during flux extraction depended on the SNR of individual images and the numbers will be mentioned when we describe the results of the data analysis and simulation.
The observation also has diffuse polarization, but, as mentioned before, in the 3C295 field this emission is small compared to the 3C196 field, motivating the use of this field rather than the latter; the diffuse emission in the real data set a lower limit on the accuracy of the measurements in the data.

\begin{figure}
\includegraphics[width=\linewidth]{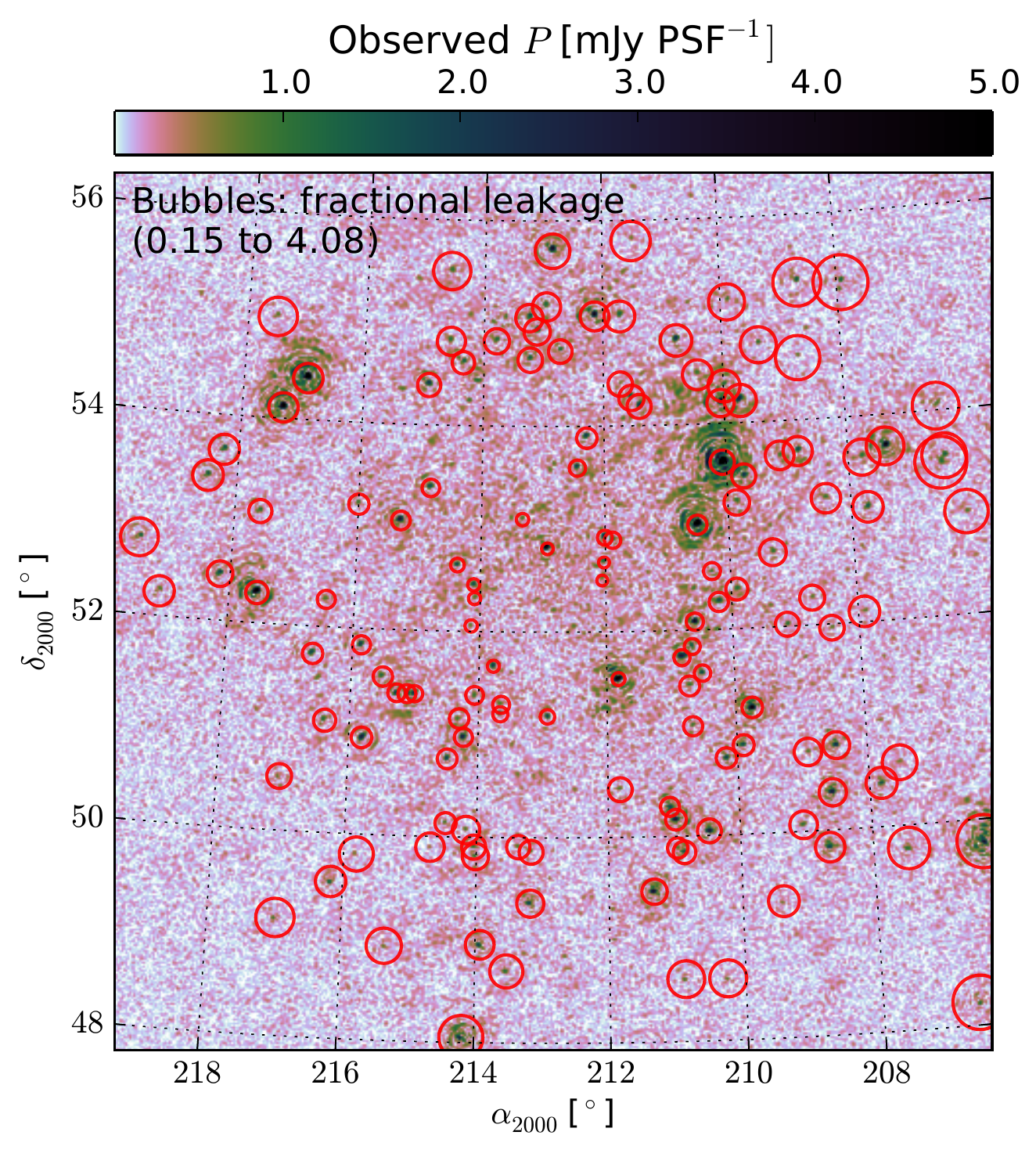}
\caption{Observed polarized emission ($|P+iQ|$) in the 3C295 field after averaging 149 frequency channels. Most of these point sources are leaked from Stokes $I$ due to instrumental polarization. Only six sources among them were found to be intrinsically polarized as shown in Fig. \ref{f:obs-intr}. The size of the bubbles indicate the amount of leakage as a percentage of Stokes $I$ flux, i. e. $m_P$ of our figures of merit.}
\label{f:obs}
\end{figure}

It should be noted that we do not include the effects of the total intensity of the diffuse foregrounds in our simulations.
One could argue that we would be affected by the total intensity of the diffuse emission here, if the emission was sufficiently bright. But, we have seen that this is not the case.
In fact, we could not detect any diffuse emission in Stokes $I$ even after removing the brightest compact sources.
More sources have to be subtracted before we can start looking for diffuse emission in total intensity.
However, the case is very different in polarization. Polarized diffuse emission can be comparable to both the instrumentally and intrinsically polarized compact sources.
But, again, we have seen that in the 3C295 field that is not the case.
In fact, less contamination from diffuse polarized emission was the very reason we chose this field for this experiment.
For example, note in Fig. \ref{f:obs} that very few compact sources are seen through diffuse polarized emission, and even the sources that are, are much brighter than the diffuse emission around them.
More details about this figure are described below in section \ref{s:obs}.

\subsection{Figures of merit} \label{s:fom}
The figures of merit used in this paper are mainly the fractional linear polarization leakages, equivalent to the degrees of polarization.
However, we also calculate the leakage for Stokes $Q,U$ separately.
The following three parameters are most frequently used:
\begin{equation}
m_P = \frac{|Q+iU|}{I}\times 100
\end{equation}
\begin{equation}
m_Q = \frac{Q}{I} \times 100
\end{equation}
\begin{equation}
m_U = \frac{U}{I} \times 100.
\end{equation}
From now on, $m_P$, $m_Q$ and $m_U$ will refer to the the observed data, and for the simulated data we will use $m'_P$, $m'_Q$ and $m'_U$, the ratios of the corresponding simulated Stokes parameters. The ratio parameters $m_P/m'_P$, $m_Q/m'_Q$, $m_U/m'_U$ and the difference parameters $m_P-m'_P$, $m_Q-m'_Q$ and $m_U-m'_U$ are the figures of merit we are most interested in.
$m_P$ of the sources should follow a Rice distribution as they are essentially the degrees of polarization \citep[for a review see][]{tr14}.

Note that these parameters are different from the IXR$_M$ introduced in section 2.
$m_P$ is most closely related to the IXR$_M$ and, as Stokes $V$ leakage is $\sim 3$ orders of magnitude lower than the linear polarization leakage (e. g. see figs. 6 and 11 of A15), the value of $m_P$ should be comparable to IXR$_M$ in magnitude.
However, we would like to point out that IXR$_M$ is calculated directly form the model of the beam, whereas $m_P$ is calculated either from the observed data or from the data created by applying the model beam on the sky and, also, in this case the data is averaged over 8 hours within which time the sky moves in the beam.
In case of instrumental polarization, $m_Q$ is determined by $M_{10}$, $m_U$ by $M_{20}$ and $m_P$ by a combination of them.

\subsection{Rotation Measure Synthesis}
A good way to distinguish between the intrinsic and the instrumental polarization is rotation measure (RM) synthesis \citep{br}.
A linearly polarized wave can undergo Faraday rotation, the rotation of its polarization angle ($\chi$), during its journey from the source to the observer if there are magnetized plasma in between.
This wavelength-dependent rotation is quantified by rotation measure, defined as $d\chi/d\lambda^2$, which is equivalent to Faraday depth
\begin{equation}
\Phi = 0.81 \int_{\rm source}^{\rm observer} n_e B_\parallel dl
\end{equation}
if the intervening magneto-ionized medium is assumed to be a single screen; here $n_e$ is the density of electrons and $B_\parallel$ is the magnetic field component along the line-of-sight component $dl$.
$\Phi$ and $\lambda^2$ are a Fourier conjugate pair, and this Fourier relationship is the basis of RM-synthesis, a per-pixel one dimensional Fourier transform along $\lambda^2$ for a multi-frequency data.
If a source is intrinsically polarized, Faraday rotation will introduce spectral structures in the broadband signature of the source.
The more it fluctuates along $\lambda^2$ the higher Faraday depth it will appear at, a basic consequence of the Fourier relationship.
On the other hand if the source is not intrinsically polarized, the only broadband signatures that it will have in its polarization is that of the `differential' beam, which is very smooth along frequency (as shown in Fig. \ref{f:beamnu}), and the ionosphere.
Due to the spectral smoothness of the beam, and due to the fact that we do not apply an ionospheric RM correction, instrumental polarization will produce a strong signal at $\Phi=0$ rad m$^{-2}$.

We have performed RM-synthesis---using the code written by Michiel Brentjens\footnote{\url{https://github.com/brentjens/rm-synthesis}}---in our analysis mainly to distinguish between the intrinsic and instrumental polarization, which is necessary if we want to compare the leakage predicted by the model beam with the instrumental part of the linear polarization seen in the observed data.
\begin{figure}
\includegraphics[width=\linewidth]{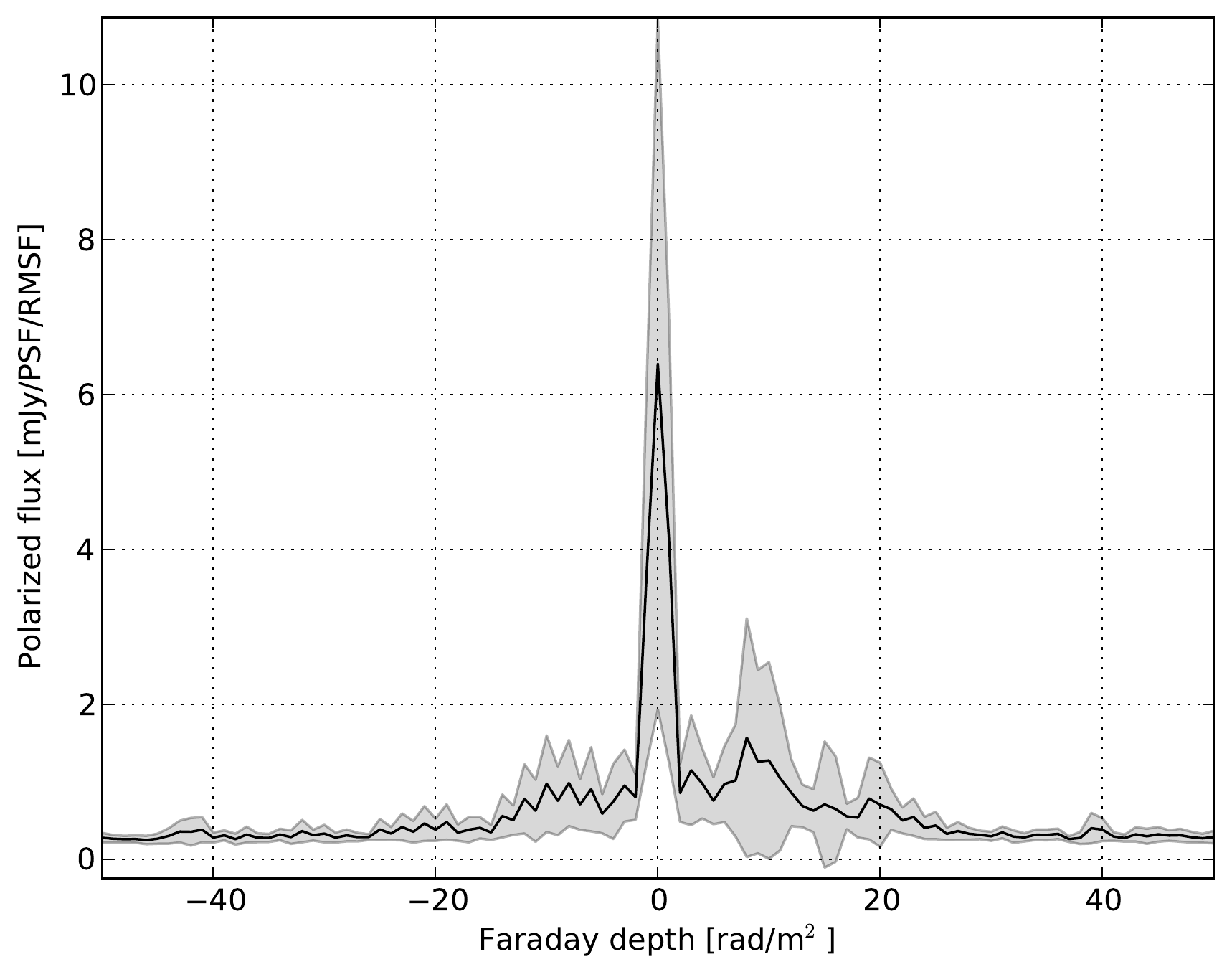}
\caption{RM profiles of the polarized point sources in the 3C295 field.
The average (solid line) and standard deviation (shaded region) of the fluxes of almost 100 compact sources have been plotted here at each Faraday depth.
What one can understand from the plot is that almost all the sources have peak flux at around $\Phi=0$ rad m$^{-2}$. 
Only 16 sources have higher-$\Phi$ peaks and even in that case most of the peaks were due to the sidelobe of the RMSF.}
\label{f:obs-rm}
\end{figure}
The image obtained after RM-synthesis is usually called Faraday dispersion function $F(\Phi)$, which is just the polarized surface brightness per unit Faraday depth.
We have not cleaned $F(\Phi)$ in our analysis, which means in our case $F(\Phi)$ is actually the polarized surface brightness convolved with the rotation measure spread function (RMSF), the equivalent of power spread function (PSF) in imaging.
However, to clearly determine the fluxes and degrees of polarization of the instrumentally polarized sources, we have subtracted the RMSF from the Faraday depth profiles, $F(\Phi)$ as a function of $\Phi$, for the sources that could be confused with sidelobes.
Sometimes a source can appear at a higher Faraday depth even if it is not intrinsically polarized due to the sidelobes of the RMSF.
However, sidelobes are usually symmetric whereas real RM structures are not.
By subtracting the RMSF we could eliminate the possibility of false detection of intrinsic polarization.

\subsection{Direction dependent calibration}
Both modeling and calibration have been or are being tested for removing polarization leakage in Fourier space.
In the former case, leakages are predicted using a model primary beam of the instrument and then deconvolved from visibilities which is essentially similar to primary beam correction in Fourier space.
One such method, called AW-projection \citep{ta}, was tested in A15, and it was found to be able to remove up to 80\% leakage.
The latter method solves for the leakages instead of modeling them by minimizing a leakage-free data set, simulated from a sky model, with the observed data toward different directions; the solutions thus produced are then applied on the observed visibilities to remove leakage.
{\tt SAGECAL} is being used as the standard tool for direction dependent calibration and source removal from Stokes $I$ in the LOFAR-EoR key project.
It can also be used to remove point sources from polarized data \citep{je15}.
Like self-calibration, {\tt SAGECAL} tries to solve for gains to match the model visibilities with the observed ones, but {\tt SAGECAL} does it for all given directions instead of just one.
If the solutions are good, the corrected data after multiplying the inverse of the gains with the observed visibilities should correspond very closely to the model visibilities.
If the model visibilities are calculated without taking into account the primary beam, without any leakage from Stokes $I$ to $Q,U$, {\tt SAGECAL} should be able to blindly incorporate the beam and the corresponding leakage terms in its gain solutions.
{\tt SAGECAL}'s performance in this regard has not been tested yet,\footnote{A special concern in this regard is the unitary ambiguity that might cause the beam-incorporated solutions to appear differently rotated than the actual beam \citep{ya12}.} and here we perform one such test.
We will show how well it can remove instrumentally polarized point sources by incorporating beam-leakage terms into the gain solutions.
For this we have run DD calibration on the DI calibrated data set and then used RM synthesis to see if {\tt SAGECAL} removed the sources at all Faraday depths.

\section{Results}

We will describe the polarization leakage found in the  observed data and the simulated observations separately and then go on to compare them to demonstrate the accuracy of the model beam.

\subsection{Observed polarization leakage} \label{s:obs}
We extracted the peak fluxes of 138 sources from the frequency-averaged $P$ image of the 3C295 field.
The faintest source in our list had a flux of 1 mJy which is $6.5\sigma$ above the noise level in the averaged image.
Most of these sources are instrumentally polarized and we could find their Stokes $I$ counterparts from which they were leaked.
After finding Stokes $I,Q,U$ fluxes of all the sources, we could calculate $m_P$, $m_Q$ and $m_U$ of the sources.
The degrees of linear polarization $m_P$ are shown by the bubble sizes in Fig. \ref{f:obs}, that ranges from 0.15\% to 4\%.
The trend of increasing $m_P$ as we go out from the phase center is also clearly visible suggesting the effect is a systematic one, and principally caused by the primary beam of the instrument; compare this with the increase of leakage as a function of distance from the phase center seen in Fig. \ref{f:ixr2}, \ref{f:ixr1}.

We then proceed to create the RM profiles, i. e. $F(\Phi)$ as a function of Faraday depth, for all sources detected in $P$.
The average and standard deviation of the fluxes of all the sources at each Faraday depth is shown in Fig. \ref{f:obs-rm}.
We can already see from this figure that the fluxes of most of the sources peak at around a single Faraday depth, and that peak is always around $\Phi=0$.
The 16 sources that show peaks at higher Faraday depths along with the 0-peak were isolated.
Among them, only 6 could be identified as intrinsically polarized, as described in section \ref{s:intr}.
In case of the other 10 sources, either their peaks were caused by the sidelobe of the RMSF, or the SNRs of the peaks were too low to be considered as a detection.
We did not take these 16 sources into account while comparing the predicted and observed leakages to calculate the accuracy of the beam model.
\begin{figure}
\includegraphics[width=\linewidth]{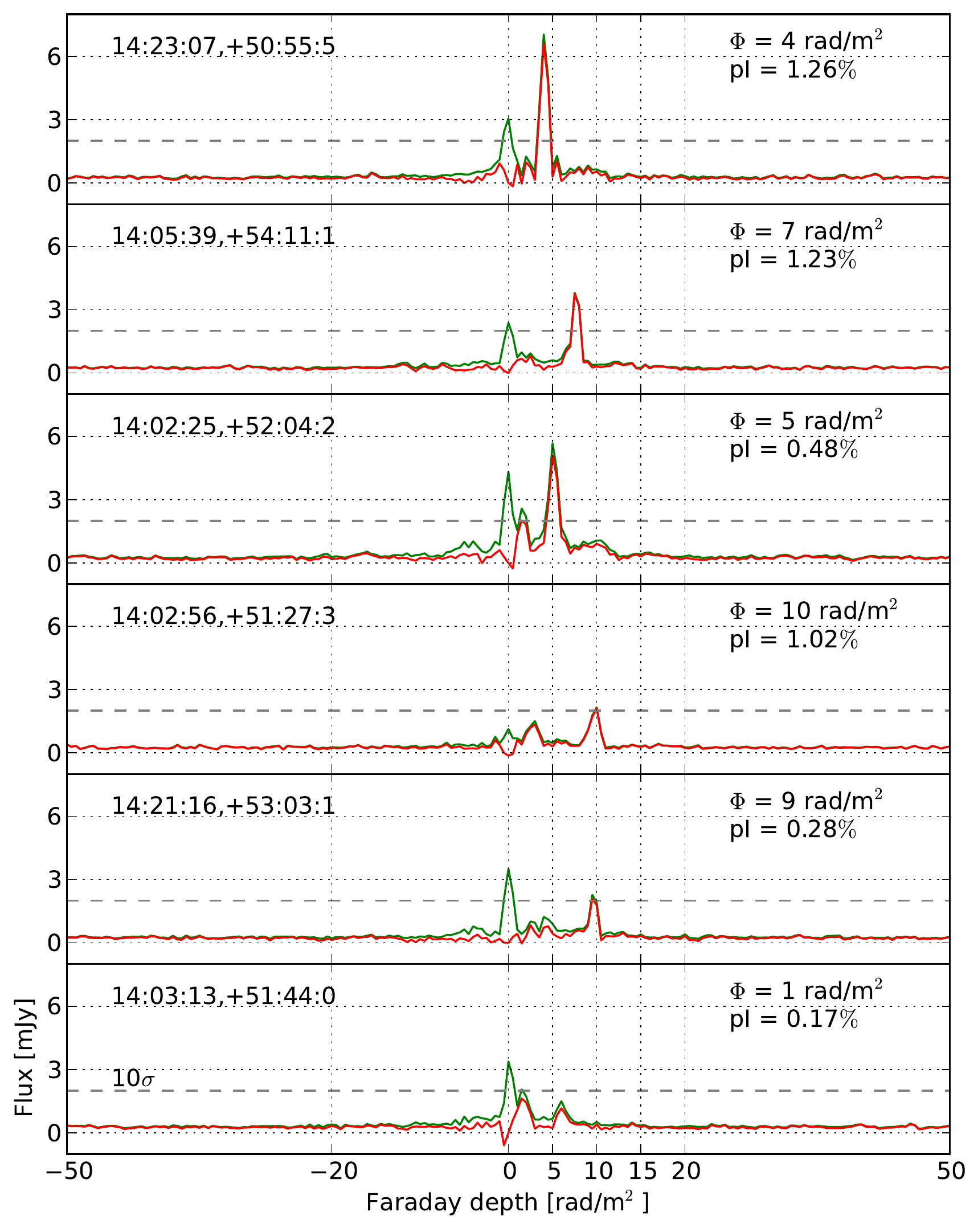}
\caption{RM profiles of the 6 intrinsically polarized point sources in the 3C295 field (in green). The red lines show the profiles after the product of the RMSF and the peak flux at $\Phi=0$ has been subtracted from the green lines. The dashed line shows the $10\sigma$ level. The positions of the sources in RA, DEC, and their Faraday depths and degrees of polarization are shown in each panel. Here, the resolution (FWHM) in Faraday depth is 1 rad m$^{-2}$.}
\label{f:obs-intr}
\end{figure}

\subsubsection{Intrinsically polarized point sources} \label{s:intr}
We have found 6 intrinsically polarized compact sources in the 3C295 field.
The RM profiles of these sources are plotted in Fig. \ref{f:obs-intr}.
The `dirty' RM profiles (convolved with the RMSF) are shown in green, and the red line shows only higher RM peaks as it was created by subtracting the product of the RMSF and the 0-peak from the `dirty' profile.
Only three sources show more than 1\% polarization and the minimum degree of polarization is only 0.17\%.
A comparison of the green and red curves in Fig. \ref{f:obs-intr} shows that for most of the sources intrinsic polarization is either comparable to or more than the polarized flux seen at around $\Phi=0$.
It should be noted that the Faraday depth and the polarization fraction are affected by the ionospheric Faraday rotation and the depth and beam depolarizations.
As we did not correct for these effects, the measured values (shown on the top right corners of each panel in Fig. \ref{f:obs-intr}) cannot be considered to be the true values.
But knowing the true values is not necessary for calculating the accuracy of the beam model; knowing whether the sources are intrinsically polarized or not is enough.

\begin{figure}
\includegraphics[width=\linewidth]{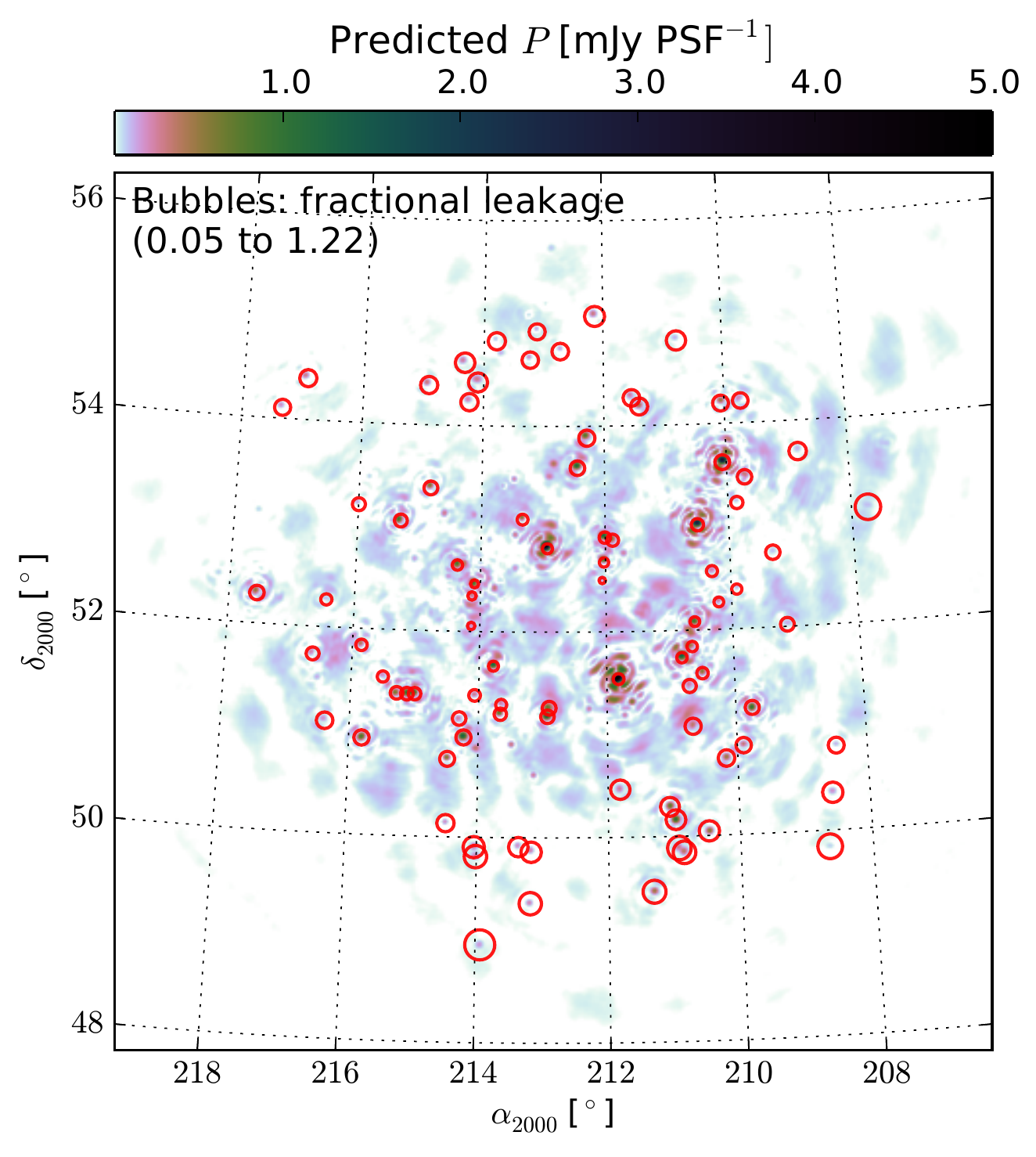}
\caption{Predicted polarization leakage for the point sources in the sky model created from observation. Sources outside the FoV are not visible because the model beam is much more attenuated than the real beam outside its FWHM. The size of the bubbles represent the fractional leakages as a percentage of the Stokes I flux.}
\label{f:model}
\end{figure}
\begin{figure}
\centering
\includegraphics[width=\linewidth]{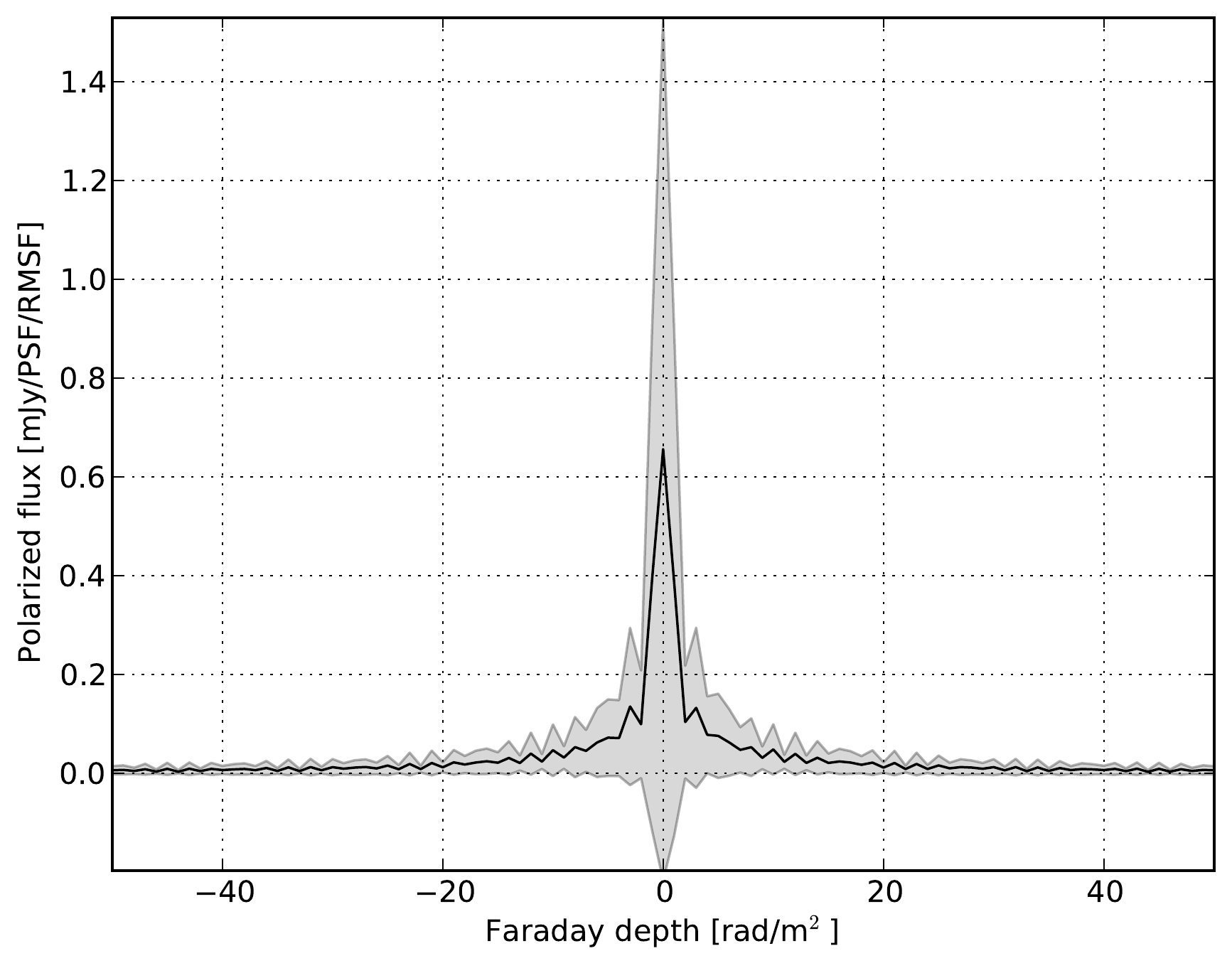}
\caption{RM profiles of the simulated polarized point sources in the 3C295 field.
The average (solid line) and standard deviation (shaded region) of the fluxes of almost 100 compact sources have been plotted here at each Faraday depth.
In contrast to Fig. \ref{f:obs-rm}, here all sources have their peak flux at around $\Phi=0$ rad m$^{-2}$.}
\label{f:model-rm}
\end{figure}
\begin{figure}
\includegraphics[width=\linewidth]{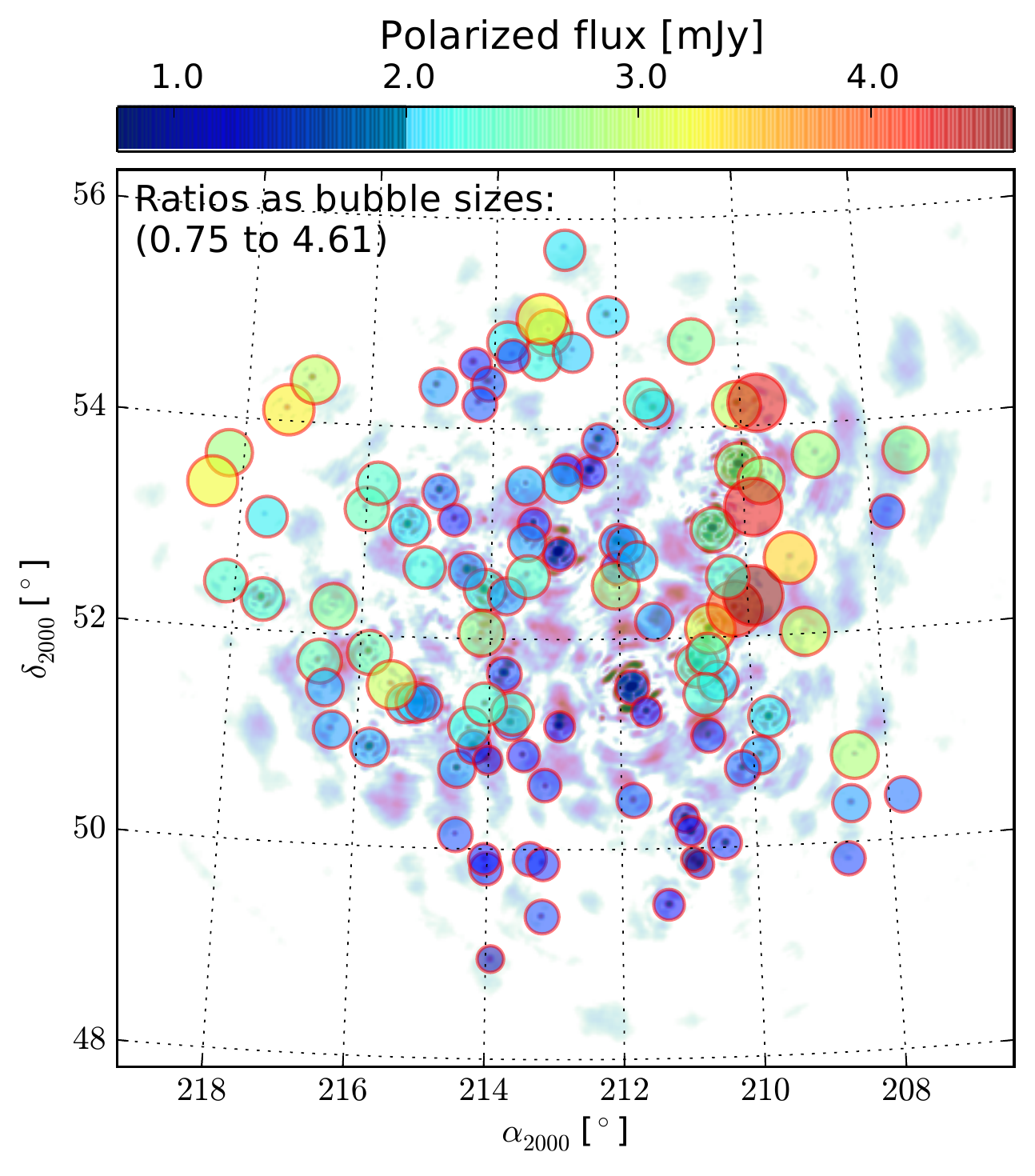}
\caption{Ratios of the observed and predicted leakages $m_P/m'_P$ represented by the size of the bubbles. It seems that for a few source model is a factor of 3--4 off from the reality, but as discussed in the text, the scenario is not that pessimistic for most of the sources, and this anomaly could be attributed to the bias and diffuse emission in the observed data. The background image is that of the simulated frequency-averaged linearly polarized image, and the color in the bubbles correspond to the polarized flux in mJy.}
\label{f:ratio}
\end{figure}

Based on previous observations, \citet{be13} stated that "one would expect to have one polarized source every four square degrees with an average polarization fraction of a few percent" between 1.4 GHz and 350 MHz.
In their 2400 deg$^2$ survey performed using MWA at 189 MHz with an angular resolution of 15.6 arcmin and a noise level of 15 mJy beam$^{-1}$, they found only one polarized point source that shows a 320 mJy peak at RM $\sim$ +34.7 rad m$^{-2}$ and a polarization fraction of $\sim$1.8\%.
On the other hand, in our 10 deg$^2$ LOFAR image averaged over 134 to 166 MHz with an angular resolution of 3.44 arcmin and a noise level of 0.15 mJy beam$^{-1}$, we have found 6 intrinsically polarized point sources.
This discrepancy is mainly due to the different sensitivities of the two observations.
A polarized source in the MWA observation had to have a flux of at least 75 mJy ($5\sigma$ above their noise level) to be considered a detection, whereas in our case even the brightest intrinsically polarized point source have a flux of only $\sim$7 mJy.
\citet{be13} did not find any polarization in the 137 point sources brighter than 4 Jy, and concluded that if any of them were polarized, the polarization fraction would be less than $\sim$2\%.
Our result is in general agreement with this conclusion, as we see that even for fainter sources---all our point sources except one are fainter than 2.5 Jy in Stokes $I$---the polarization fraction is not more than $\sim$1.3\%.

\subsection{Predicted polarization leakage}
We have identified 95 instrumentally polarized sources in the frequency-averaged image of the visiblities predicted using the unpolarized sky model created from observation.
Sources appear in Stokes $Q$ and $U$ because of the primary beam induced leakage and their degrees of polarization $m'_P$ are shown as bubbles in Fig. \ref{f:model}.
Only the sources within the first null of the primary beam are shown here, as the current software for simulating visibilities can reproduce the effects of the `real' primary beam well only within the FoV.
The polarization leakage from outside the FoV only comes in via the sidelobes and is a very small effect and because the EoR analysis is limited to the FoV, this is the only region of interest in terms of polarization leakage.
Leakages from Stokes $I$ into polarization increases as a function of distance from the phase center and they range from 0.05 to 1.22 per cent as shown by the sizes of the bubbles.
This is consistent with the fractional leakages observed in the 3C196 field which is expected as the two fields roughly have similar declinations.
The average and standard deviation of the fluxes of all the sources are plotted at each Faraday depth in Fig. \ref{f:model-rm} and the contrast with Fig. \ref{f:obs-rm} is clearly visible.
In the previous figure, intrinsically polarized sources were found at higher Faraday depths due to the rotation of their polarization angle by intervening magneto-ionic medium, but in the latter figure there are no such sources as here the polarization is caused only by the spectrally smooth primary beam.
The sources that do appear at slightly higher Faraday depths than 0 rad m$^{-2}$ in the latter figure do so only due to the sidelobe of the RMSF.

\begin{figure}
\includegraphics[width=\linewidth]{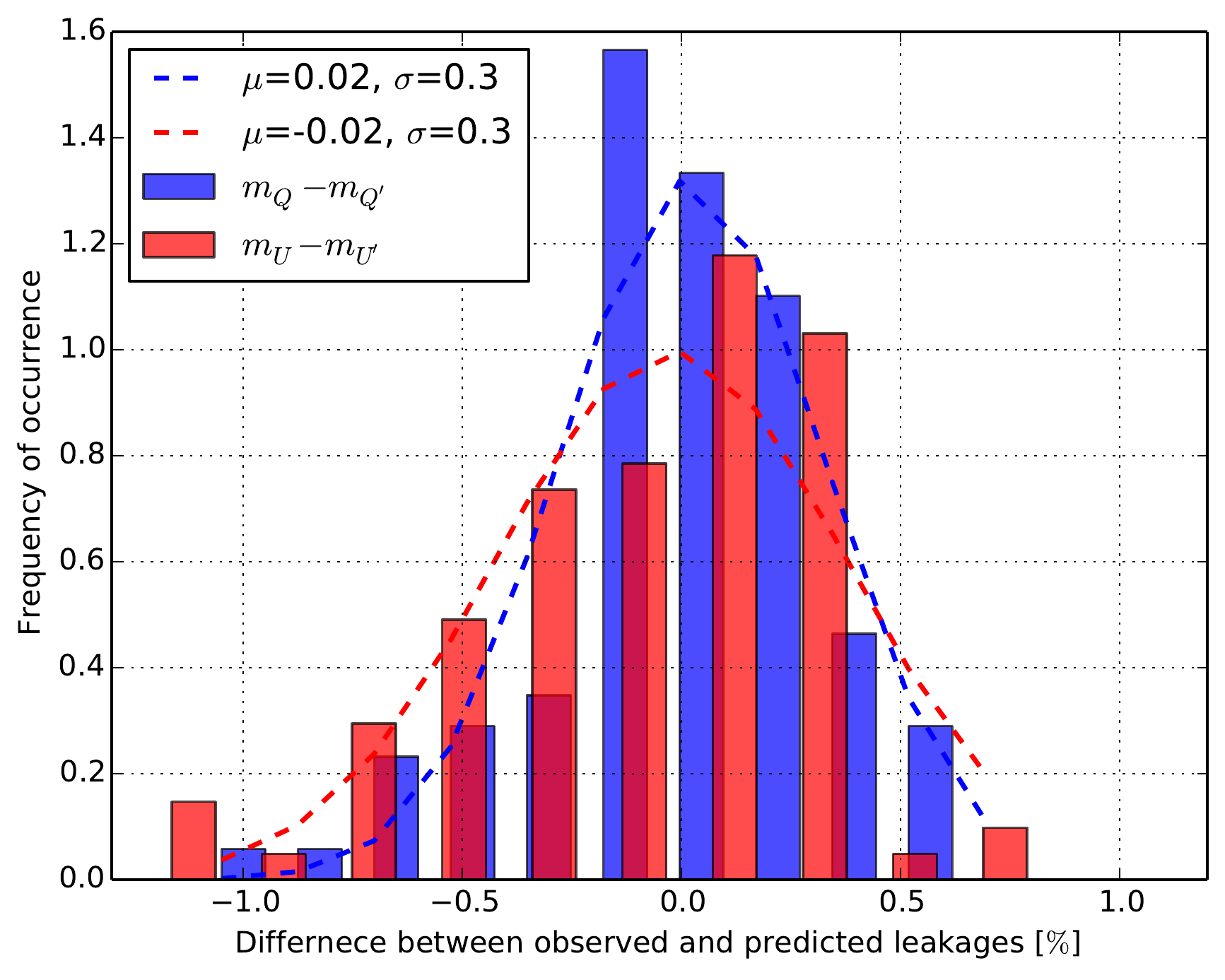}
\caption{Difference between the fractional observed and predicted leakages into Stokes $Q$ (blue) and $U$ (red). Both of them follow approximately a Gaussian with means close to zero and a standard deviation of 0.3. The dashed lines show the Gaussian fits to the bar chart.}
\label{f:hist}
\end{figure}

\subsection{Accuracy of the beam model}
As a first step toward understanding the accuracy of the primary beam model, we have compared the degrees of polarization of the observed and predicted polarized sources, i. e. $m_P$ and $m'_P$, by taking their ratio.
The parameter $m_P/m'_P$ for the sources found in both the observed and the predicted images is plotted in Fig. \ref{f:ratio}.
Both the size and color of the bubbles correspond to the ratios of the observed and predicted leakages.
The most general trend in Fig. \ref{f:obs}, \ref{f:model} and \ref{f:ratio} is that the observed leakage is almost always more than the one induced by the model beam which seems to show that the model beam is under-predicting the leakage, but one should note that the observed data has noise and diffuse emission that contribute to the estimation of source fluxes.
The observed leakage is seen to be 0.75 to 4.61 times higher than the predicted leakage, but for most of the sources the ratio is less than 2.
The overestimation of the degrees of polarization in the observed data could be due to the well-known bias in the presence of noise \citep{ss85}, and the diffuse emission faintly visible in Fig. \ref{f:obs}.

\begin{figure}
\includegraphics[width=\linewidth]{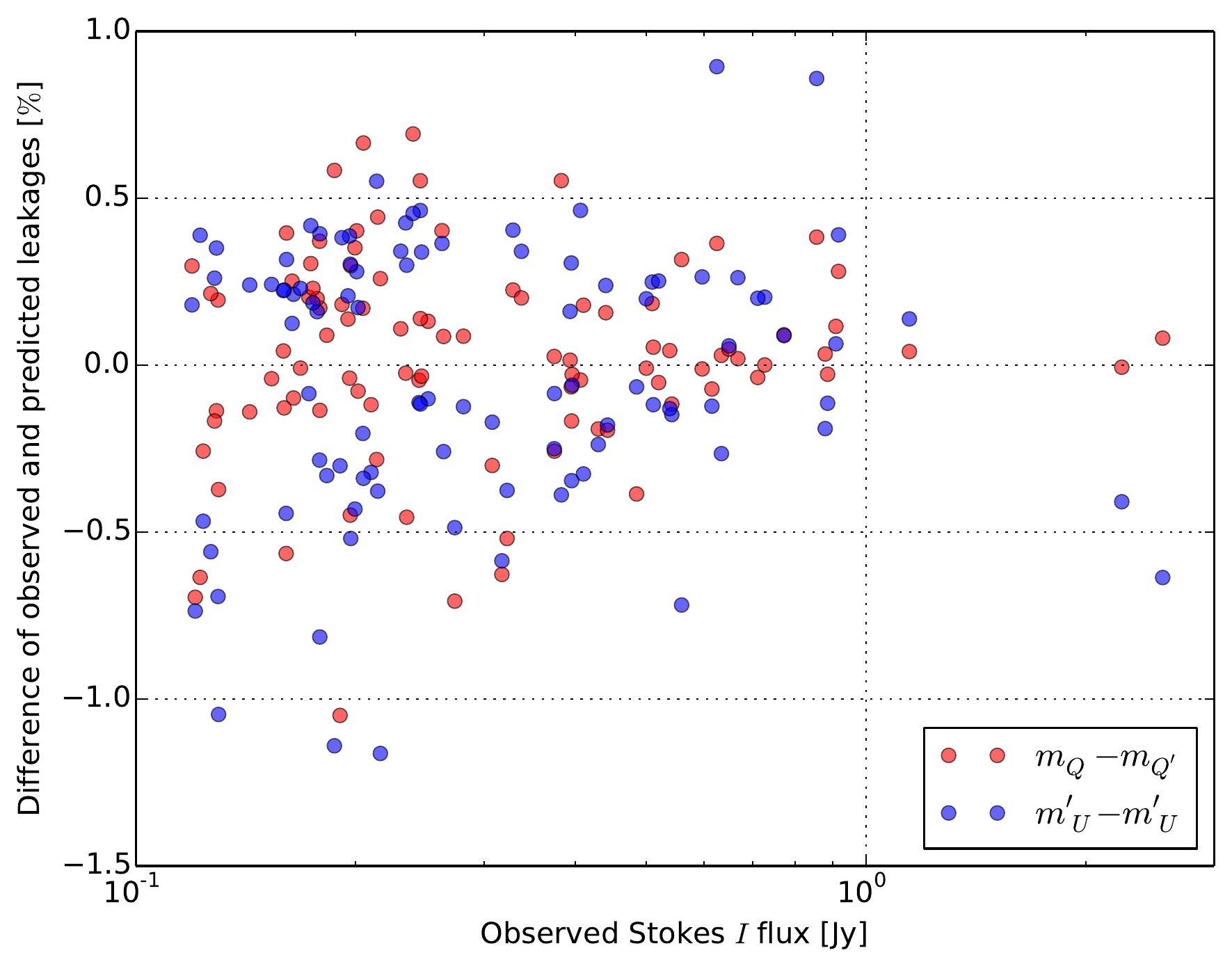}
\caption{Difference of observed and predicted leakages as a function of the corresponding Stokes $I$ fluxes of the sources. More scatter at the dimmer end indicates errors related to extracting flux of dim sources.}
\label{f:mqmu-I}
\end{figure}
\begin{figure}
\includegraphics[width=\linewidth]{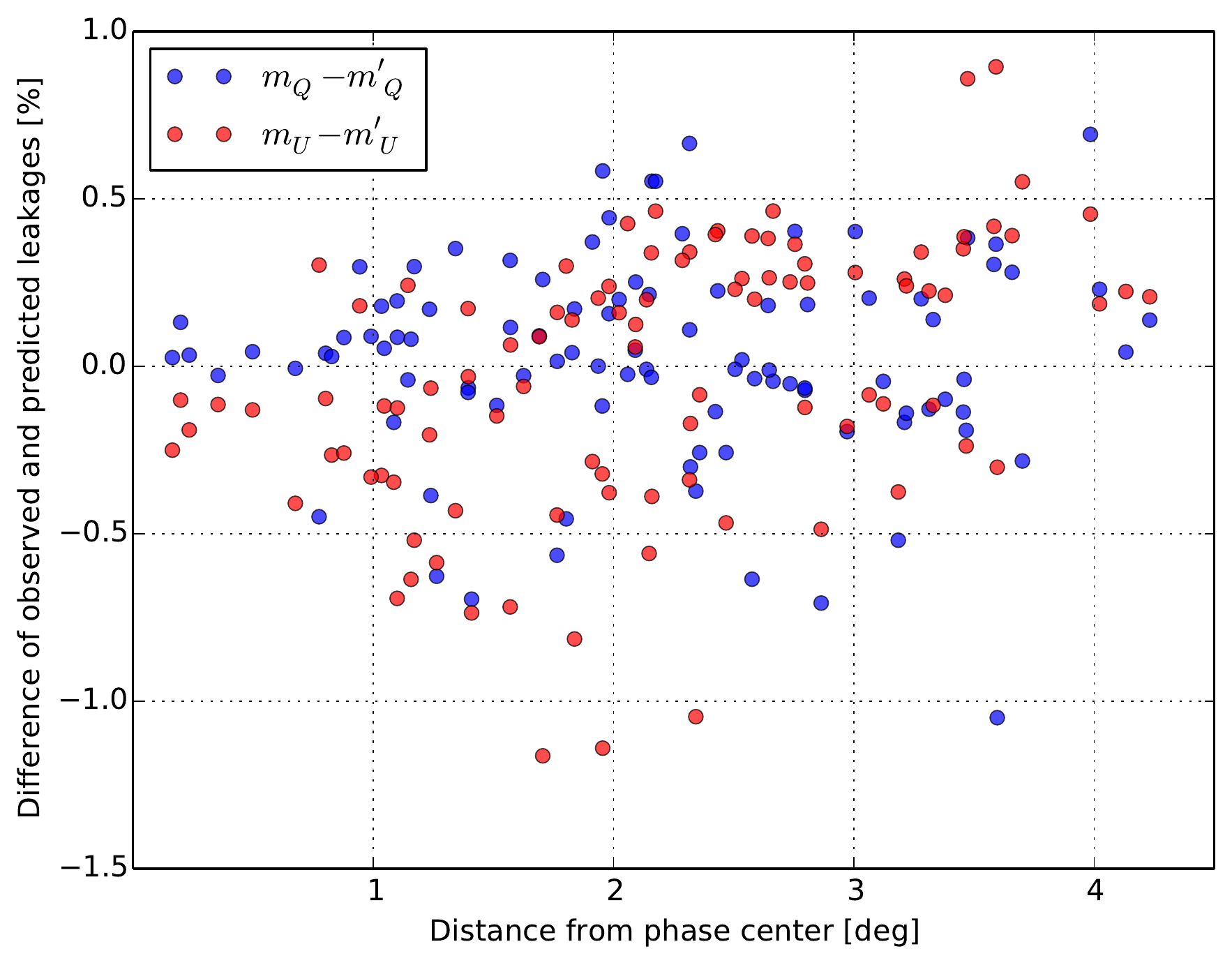}
\caption{The differences between the observed and predicted leakages for different sources are plotted against the corresponding distances from the phase center for both Stokes $Q$ (blue) and $U$ (red).}
\label{f:mqmu-d}
\end{figure}

A more natural way to calculate the accuracy of the beam would be to compare the leakages into Stokes $Q$ and $U$ separately and take the difference between the observed and predicted leakages, i. e. $m_Q-m'_Q$ and $m_U-m'_U$, instead of their ratios.
A bar chart of these difference parameters are plotted together in Fig. \ref{f:hist}.
As individual Stokes parameters follow Gaussian noise statistics, their difference should also be Gaussian, and although here we rescale the Gaussian by taking the difference between the \textit{ratios} of Stokes parameters and although the diffuse foreground might not follow Gaussian noise, the distribution still approximately follows a Gaussian.
Both $m_Q-m'_Q$ and $m_U-m'_U$ follow approximately a Gaussian with means close to zero (0.02 for $Q$ and -0.03 for $U$) and a standard deviation $\sigma$ of 0.3.
Therefore, we can say with a 68\% certainty that the leakage predicted by the model beam of LOFAR will be 30\% different from the actual leakage. If the actual leakage is $\sim$ 1\%, the model beam might predict the leakage to be around 0.7\% -- 1.3\%.

We have calculated the uncertainty in the prediction of the beam model induced polarization leakage, but there are uncertainties in that uncertainty arising from the errors in extracting fluxes of the sources.
To show these uncertainties we plot the FoM $m_Q-m'_Q$ and $m_U-m'_U$ for the sources as a function of their Stokes $I$ fluxes in Fig. \ref{f:mqmu-I}, and as a function of their distances from the phase center in Fig. \ref{f:mqmu-d}.
In the former plot, as the flux of the source decreases thereby decreasing the local SNR of the source and enhancing the effect of the Gaussian noise, the random scatter of the aforementioned FoM increases.
Therefore, this trend can be attributed to the Gaussian noise in the image that leaves its imprint on the extracted fluxes.
As sources are attenuated as we go away from the phase center due to the azimuthally decreasing primary beam, we should expect an increase in the scatter of the FoM as we go outward from the phase center, and this is exactly what we see in the latter figure.
Hence, this incremental trend of the FoM as a function of distance from the center should not be attributed to a systematic bias in the model of the beam, but again to the imprint of the image noise on the extracted fluxes.

As, here, we are mainly limited by the image noise and the errors in extracting fluxes of faint sources, one would expect the uncertainty in the calculation of the accuracy of the beam model to go down if a higher flux density cut is used.
And we see exactly this trend.
We have taken only the 18 sources brighter than 600 mJy in Stokes $I$ and made bar charts similar to that of Fig. \ref{f:hist} and found that the standard deviation indeed improves significantly---although the mean of $m_Q-m'_Q$ remained 0.02\%, its standard deviation improved to 0.1\%.
On the other hand, for the 26 sources brighter than 500 mJy, the $\sigma$ was found to be 0.2\% showing that, due to the effect of the image noise, $\sigma$ increases as we include more fainter sources.
The contribution of flux extraction error in the calculation of the accuracy of the model beam can also be seen clearly by comparing Fig. \ref{f:hist} and \ref{f:mqmu-I}---the sources for which the difference between the observation and prediction is more than 0.5\% are the ones with low flux and high scatter, and if we discard these sources the bar chart becomes narrower and exhibits a lower standard deviation for both Stokes $Q$ and $U$.
Note that the standard deviation in these figures is contributed solely by the observed images as there was no additive noise in our simulation.
The 18 brightest sources provide a clean model that is precise enough to predict leakage more accurately over the FWHM of the primary beam.
Therefore, we can now say that the errors are $\le 10\%$ on the predicted levels of leakage of $\sim 1\%$ typically in $68\%$ of the cases, i. e. in these cases polarization leakage after calibration with the nominal LOFAR beam should be $\le 10^{-3}$ of Stokes $I$ within the FoV.

\subsection{Direction dependent calibration}
\begin{figure}
\includegraphics[width=\linewidth]{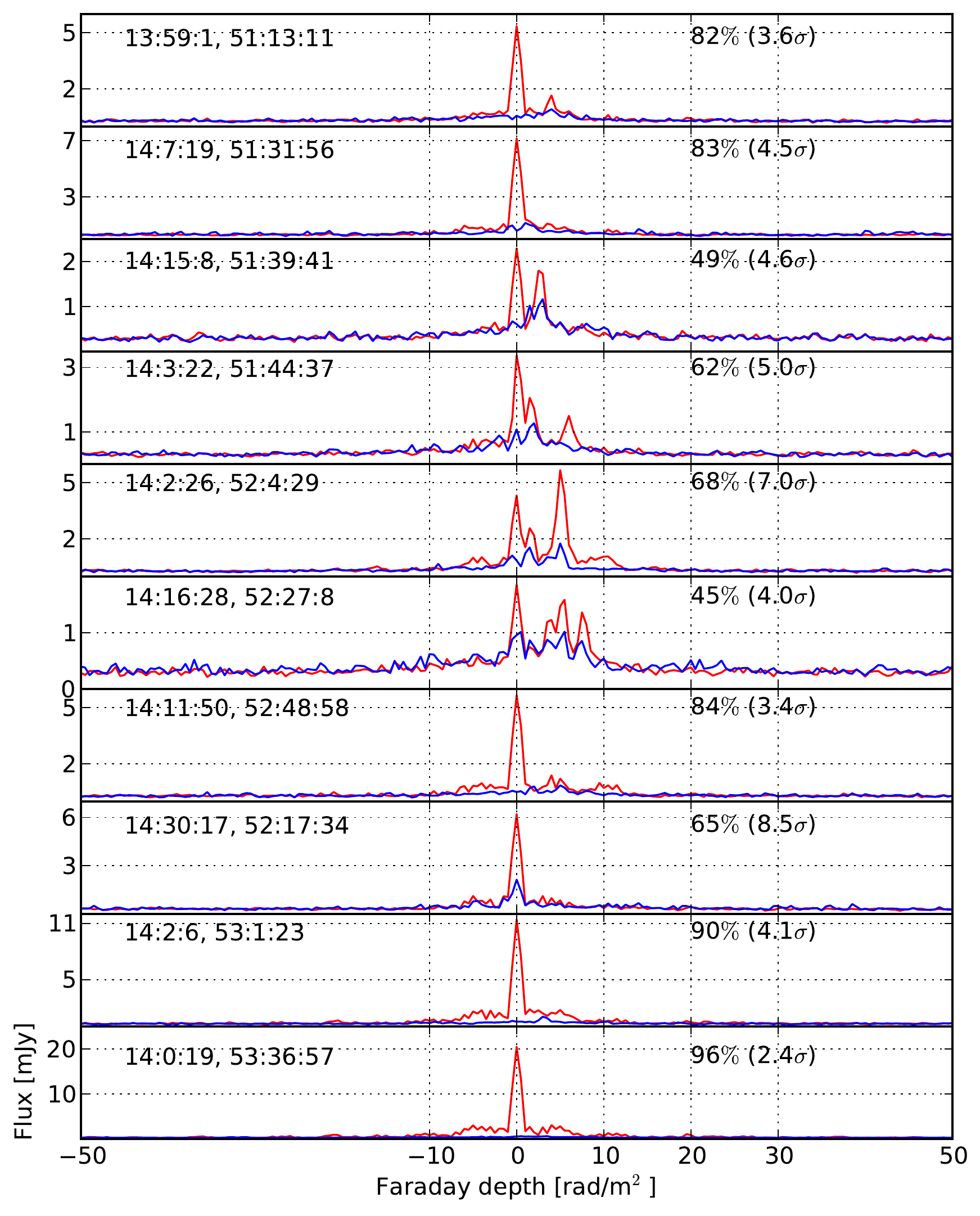}
\caption{RM profiles of the 10 sources used in DD calibration both before (red) and after (blue) the calibration. For most of the sources, more than 80\% flux could be subtracted using this calibration. RA and DEC of the sources are given on the top left corner of every panel for ease of comparison. The texts on the top right corners show percentages of flux subtracted, and the residual levels with respect to the image noise (in brackets).}
\label{f:sage}
\end{figure}

We solved for DD gains toward 10 clusters using {\tt SAGECAL} \citep{ka11,ka13}.
Instead of solving toward the direction of every source in the sky model, SAGECAL groups the sources into different clusters and solves for the gains toward the center of each cluster \citep{ya13}.
However, in our case, each cluster had only one source, as solving for only the brightest sources is sufficient for our demonstration purpose.
As the sky model was completely unpolarized, SAGECAL should subtract polarized flux at all Faraday depths irrespective of instrumental or intrinsic polarization.
RM profiles of the 10 sources were created both before and after SAGECAL and they are shown together in Fig. \ref{f:sage} in red and blue respectively.
The figure shows that the sources with high SNR, i. e. half of the sources, were subtracted to more than 80\%, and the brightest two sources were subtracted to $\ge 90\%$.
Local noise level in these images was on average 0.2 mJy, and the brighter sources were removed sufficiently close to the noise level.
The brighter the source, the better it was removed; the residual of the 20 mJy source was only 2.4 $\sigma$ above the local noise.
Residuals after subtracting all the sources are mentioned on the top right corner of each panel, and we see that most residuals are $< 5\sigma$.

\section{Discussion and Conclusions}

We have calculated the accuracy of the nominal model beam of LOFAR---created from the EM simulations of the ASTRON antenna group \citep{ha11}---by comparing the leakages predicted by the model beam with that of the observation of the 3C295 field.
Fig. \ref{f:beam} shows the model beam of a typical station (left panel), and the mismatch between the beams of the two dipoles (right panel), and Fig. \ref{f:beamnu} shows that the position of the sidelobe of the beam varies smoothly along frequency.
Although the mismatch of the feed-beams already shows the extent of the polarization leakage, we have quantified the polarimetric performance of the beam using the IXR$_M$, the Mueller matrix version of the intrinsic cross-polarization ratio, a standard figure of merit for measuring the polarimetric performance of low-frequency arrays \citep[see, e. g.,][]{dla15}.
Fig. \ref{f:ixr2} and \ref{f:ixr1} show that the polarimetric performance of low-frequency aperture arrays like LOFAR is best near the phase center of the field and when the field is close to its culmination point.
However, narrowing the field of view or filtering out the observations close to horizon result in reduced sensitivity and a balance between data filtering and calibration and modeling of the systematic errors needs to be maintained.
In A15, we showed that taking data only within the central 3 degrees decreases the effect of polarization leakage.
Here, from Fig. \ref{f:ixr1}, we see the significant improvement of polarimetric performance close to the zenith, and further work is needed to establish a balance between the calibration and/or modeling of the DD systematic effects and the avoidance of the systematics dominated observation.
Note that we did not use the IXR$_M$ directly while calculating the accuracy of the model beam, but the figures of merit we used for this purpose is very closely related to IXR$_M$, as explained in section \ref{s:fom}.

The prediction of polarization leakage in the `EoR window' of the cylindrical PS can be made more robust in the context of LOFAR based on the calculations of this paper.
A15 found that even without any leakage correction the simulated EoR signal is higher than the rms of the leakage in a significant portion of the cylindrical PS, and this EoR window extends to almost the whole instrumental $k$-space of LOFAR if 70\% of the leakage could be removed.
In the current paper, by comparing the leakages from Stokes $I$ to $Q,U$, we have found that the prediction of the beam, in $68\%$ of the cases, will have an error of $\le 10\%$, i. e. if the predicted leakage is 1\%, the actual leakage might be between 0.9\% to 1.1\%.
Therefore, if the differential beam effects are taken out perfectly using the nominal model beam of LOFAR, the errors in the correction will be $\le 10\%$, i. e. the residual leakage in Stokes $Q,U$ will be $10^{-3}$ of Stokes $I$ flux.

We could calculate the accuracy of the beam model only up to the first null; accuracy of the sidelobes of the model could not be calculated for two interconnected reasons.
First, the beam model under-predicts leakage on the sidelobes to some extent which can be seen by comparing the observed (Fig. \ref{f:obs}) and the simulated (Fig. \ref{f:model}) images.
In the former figure, some sources can be seen on the sidelobes, whereas in the latter all sources are within the FoV (note that the FoV would also change with frequency).
Of course, the accuracy of the model beam could still be calculated, if we could quantify the under-prediction, and that's where the second reason comes in.
The Stokes I fluxes of the sources in the sidelobes were already very low as they were attenuated by the primary beam, and when we predicted leakage from these "faint" sources, the resulting leakage was even lower.
So, we could not find compact sources bright enough to give rise to a detectable polarization leakage, even after the under-prediction of the beam, that would make the calculation of the accuracy possible at these distances from the phase center.
Due to this limitation, we claim our measurement of the accuracy of the beam model to be reliable only within the FoV.
However, a future paper in this series (in preparation) will take into account both the leakage and the accuracy of the beam model farther away from the phase center, as they are crucial for EoR experiments.

The result of this experiment obtained using the $I\rightarrow Q,U$ leakages should hold true even for the $Q,U\rightarrow I$ leakages, as their relationship is symmetric for both the on-axis \citep{sa96} and off-axis (e. g. see fig. 2b of A15) beams.
Therefore, we can say that the beam model used to predict the $Q,U\rightarrow I$ leakage in A15 had a 10\% error, and if the leakage could be removed, this error would be one of the constituents of the residual.
However, we do not know how well this subtraction can be performed given that the leakage is even below the noise level, let alone the total intensity of the diffuse foregrounds.
One should be careful about the uniqueness of each field in terms of both the projection effects of the beam and the diffuse polarization structure.
For example, the diffuse polarized emission in the 3C295 field is very different in both amplitude and spatial and Faraday structure from that of the 3C196 field, but the projection of the beam toward these fields are not that different as they are situated at similar declinations.

We used DD-calibration to remove leakages of compact sources from Stokes $I$ to $Q,U$ and found that for sources with sufficiently high SNR, more than 80\% of the flux could be removed and the residuals were generally very close to the local noise level.
More work is needed to see how this blind correction of leakage compare with the correction using model beam.
A good way to compare the modeling and DD-calibration approaches would be to test the effectiveness of AW-projection (using, e. g., {\tt AWIMAGER}) and {\tt SAGECAL} in removing linear polarization leakage.
However, both {\tt AWIMAGER} and {\tt SAGECAL} can remove only the leakages of compact sources from Stokes $I$ to $Q,U$, whereas for the EoR project we are interested in the leakages of diffuse emission from Stokes $Q,U$ to $I$.
Work is underway to test the effectiveness of removing this leakage using an RM-model of the diffuse emission and the nominal model beam of LOFAR.

\section*{Acknowledgments}
KMBA, LVEK, and AG acknowledge the financial support from the European Research Council under ERC-Starting Grant FIRSTLIGHT -- 258942.
VJ acknowledges the financial support from The Netherlands Organization for Scientific Research (NWO) under VENI grant -- 639.041.336.
AGdB, MM, SY, and VNP acknowledge support from the European Research Council under grant 399743 (LOFARCORE).
SZ acknowledges the support from Lady Davis Foundation and NWO VICI grant.
ITI was supported by the Science and Technology Facilities Council [grant numbers ST/F002858/1, ST/I000976/1 and ST/L000652/1].

\footnotesize{
\bibliographystyle{mn2e}

}

\end{document}